\documentclass[twocolumn]{aastex63}

\usepackage{bm}
\usepackage{amsmath}
\usepackage{color}
\usepackage{graphicx}
\usepackage{ulem}
\received{June 1, 2019}
\revised{January 10, 2019}
\accepted{\today}
\submitjournal{ApJ}

\shorttitle{Dusty/dust-free outflows in AGN}
\shortauthors{Kudoh et al.}

\begin{document}

\title{
Multiphase Gas Nature in the Sub-parsec Region of the Active Galactic Nuclei I: Dynamical Structures of Dusty and Dust-free Outflow 
}

\correspondingauthor{Kudoh Yuki}
\email{yuki.kudoh@astrophysics.jp}

\author[0000-0003-0548-1766]{Yuki Kudoh}
\affiliation{
Astronomical Institute, Tohoku University,\\
6-3 Sendai, Miyagi 980-8578, Japan}
\affiliation{
Graduate School of Science and Engineering, Kagoshima University,\\
1-21-35 Korimoto, Kagoshima 890-0065, Japan}
\affiliation{
National Astronomical Observatory of Japan, 2-21-1 Mitaka, Tokyo 181-8588, Japan}

\author[0000-0002-8779-8486]{Keiichi Wada}
\affiliation{
Graduate School of Science and Engineering, Kagoshima University,\\
1-21-35 Korimoto, Kagoshima 890-0065, Japan}

\author[0000-0003-2535-5513]{Nozomu Kawakatu}
\affiliation{
Faculty of Natural Sciences, National Institute of Technology, Kure College, \\
2-2-11 Agaminami, Kure, Hiroshima 737-8506, Japan}

\author[0000-0002-6236-5270]{Mariko Nomura}
\affiliation{
Graduate School of Science and Technology, Hirosaki University, \\
3 Hirosaki, Aomori 036-8561, Japan}
\affiliation{
Faculty of Natural Sciences, National Institute of Technology, Kure College, \\
2-2-11 Agaminami, Kure, Hiroshima 737-8506, Japan}


\begin{abstract}
We investigated dusty and dust-free gas dynamics for a radiation-driven sub-parsec-scale outflow in an active galactic nucleus (AGN) associated with a supermassive black hole $10^7 M_\odot$ and bolometric luminosity $10^{44}$ erg s$^{-1}$ based on the two-dimensional radiation-hydrodynamic simulations. 
A radiation-driven ``lotus-like'' multi-shell outflow is launched from the inner part ($r \lesssim 0.04$ pc) of the geometrically thin disk,
and it repeatedly and steadily produces shocks as mass accretion continues through the disk to the center.
The shape of the dust sublimation radius is not spherical and depends on the angle ($\theta$) from the disk plane, reflecting the nonspherical radiation field and nonuniform dust-free gas. 
Moreover, we found that the sublimation radius of $\theta \sim 20$--$60$ deg varies on a timescale of several years. 
The ``inflow-induced outflow" contributes the obscuration of the nucleus in the sub-parsec region. The column density of the dust-free gas is $N_{\rm H} \gtrsim 10^{22}$ cm$^{-2}$ for $r \lesssim 0.04$ pc. 
Gases near the disk plane ($\theta \lesssim 30$ degree) can be the origin of the Compton-thick component, which was suggested by the recent X-ray observations of AGNs. 
The dusty outflow from the sub-parsec region can be also a source of material for the radiation-driven fountain for a larger scale.

\end{abstract}

\keywords{hydrodynamics---radiation: dynamics---methods: numerical---galaxies: active---galaxies: nuclei}


\section{Introduction} \label{sec:intro}

The current paradigm of the structures of active galactic nuclei (AGNs) is that the broad emission line region (BLR) is surrounded by a geometrically thick dusty material, so-called  ``dusty torus"  (\citealt{1993ARA&A..31..473A, 2019ApJ...884..171H}; see also the review \citealt{2015ARA&A..53..365N}). 
Thanks to the recent progress in radio and infrared interferometric techniques, the structures of molecular tori and their inner edge regions were partially resolved using the Very Large Telescope Interferometer (VLTI: \citealt{2020A&A...635A..92G,2022A&A...663A..35I}, and see also \citealt{2018ExA....46..413H}) and the Atacama Large Millimeter/submillimeter Array (ALMA: \citealt{2018ApJ...867...48I,2019A&A...632A..61G,2021A&A...652A..98G,2019ApJ...884L..28I,2023arXiv230503993I}, and see also \citealt{2021agnf.book.....C}). 
In this paradigm, the inner edge of the dusty tori is determined by the sublimation of the dust grains illuminated by the AGN radiation \citep{1987ApJ...320..537B, 2012MNRAS.420..526M}. 
For the dust sublimation temperature ($T_{\text{sub}}$), the sublimation radius is

\begin{equation}
r_{ \text{sub} }=1.3 \left( \frac{L_{\text{UV}}}{10^{46} \text{erg s}^{-1}}  \right)^{0.5} \left( \frac{T_{\text{sub}}}{ 1500 \text{K} }    \right)^{-2.6},
\label{eq:sublimation_clsc}
\end{equation}
~\\
where $L_{\text{UV}}$ denotes the ultraviolet luminosity.
The reverberation mapping observations of nearby AGNs suggested that the dust sublimation radius is close to $r_{ \text{sub} }\propto L^{0.5}$  considering the size--luminosity relation  (\citealt{2006ApJ...639...46S, 2007A&A...476..713K, 2014ApJ...788..159K}). 
Based on the time variability of the optical and infrared bands used for reverberation measurements, several studies have examined the radial profile of dusty tori \citep{2011A&A...536A..78K,2021ApJ...912..126L}.
Time variability is reported to be several years or less \citep{2011A&A...527A.121K,2014ApJ...788..159K,2019ApJ...886..150M}.

However, the structures of the transition regions of dusty and dust-free gases near $r =r_{\text{sub}}$ still are not observationally clear.  
\citet{2015ApJ...806..127D} suggested that a geometrically thick neutral gas torus coexists with the BLR and bridges the gap with the dusty torus by analyzing the X-ray-selected sample of nearby AGNs. 
Based on the comparison of the X-ray spectral energy distribution (SED) between hydrodynamic models and observations, \citet{2021A&A...651A..58B} found that a Compton-thick material is necessary at $r < 0.1$ pc to reproduce the hard X-ray spectrum of the nearby (4.2 Mpc) type-2 Seyfert, the Circinus galaxy. 
Moreover, this necessary is confirmed by soft X-ray absorption features concerning warm absorbers \citep{2022ApJ...925...55O}.

The structure of an inner surface of the dusty torus has been theoretically studied by \cite{2010ApJ...724L.183K}, \cite{2011ApJ...737..105K,2011A&A...534A.121H}, and \cite{2017ApJ...843....3A,2020ApJ...891...26A} with respect to reverberation mapping based on a simple, static picture. 
However, dusty and dust-free gases around the AGN may not be static as suggested in torus-scale radiation-hydrodynamic simulations \citep{2016ApJ...819..115D,2016ApJ...825...67C, 2016MNRAS.460..980N, 2019ApJ...876..137W, 2020ApJ...897...26W}. 
\citet{2012ApJ...758...66W, 2015ApJ...812...82W} proposed that the radiation pressure and heating in AGN are essential to drive gas circulation in the central parsec region. 
The “radiation-driven fountain” model is consistent with the multiwavelength observations of the central parsec to 10 pc of the Circinus galaxy (\citealt{2016ApJ...828L..19W, 2018ApJ...867...48I, 2018ApJ...852...88W, 2018ApJ...867...49W, 2021ApJ...915...89U, 2022ApJ...934...25M, 2022ApJ...925...55O,2023arXiv230503993I}).

Their hydrodynamical studies have focused on the parsec-scale structure. 
Therefore, we do not understand the structures and dynamics of dusty and dust-free gases in the transition region, around $r_{\text{sub}} \sim 0.1$ pc for a supermassive black hole (SMBH) $M_{\text{SMBH}}=10^7 M_{\odot}$ and bolometric luminosity $10^{44}$ erg s$^{-1}$ (see, Equation (\ref{eq:sublimation_clsc}))
In this study, we performed two-dimensional (i.e., axisymmetric) radiation-hydrodynamic simulations that spatially resolve the dust sublimation radius to understand the structures and dynamics of the radiation-driven outflow. 
We focus on the following questions:
(1) Where and how is the radiation-driven dusty outflow launched?
(2) How do dusty and dust-free gases contribute to column density?
(3) What are the realistic shape of the dust sublimation radius and its time variability under the effect of fueling and its feedback in the AGN sub-parsec scale?

This paper is organized as follows.
In Section \ref{sec:methods}, we describe the solving equations and physical models of the dusty gas around the AGN.
The numerical method and model setup are described in Section \ref{subsec:25}.
We present the results of the radiation-driven dusty outflow for the multiscale propagation of the shell structure (\S \ref{subsec:41}). 
We also show the viewing angle dependence of the column density, dust sublimation radius, and time-variable nature of the transition region (\S \ref{subsec:42}).
The relation between the current results and the structures of the torus scale is discussed in Section \ref{sec:discussion}.
In Section \ref{sec:summary}, we summarize our findings

\section{Methods} \label{sec:methods}
Following the parsec-scale radiation-driven fountain model \citep{2012ApJ...758...66W,2015ApJ...812...82W}, we focus on the inner $10^{-4}$ pc -$1$ pc region. 
We solved the dynamics of the gas--dust disk accreting to a central SMBH ($10^7 M_\odot$)
under the effect of an anisotropic central radiation field. 
Herein, we assume an axisymmetric system to achieve high spatial resolution ($5\times10^{-4}$ pc) (\S \ref{subsec:25}).  
We also modified the dust opacity model and the source SED described in \S\S \ref{subsec:22}--\ref{subsec:24}.

\subsection{basic equations}  \label{subsec:21}

The basic equations are
\begin{eqnarray}
\frac{\partial \rho}{\partial t} + \bm{\nabla} \cdot \left[\rho \bm{v} \right]  =0,  
\label{eq:mass}
\end{eqnarray}
\begin{eqnarray}
 \displaystyle \frac{\partial \rho \bm{v}}{\partial t}
 + \bm{\nabla} \cdot \left[ \rho  \mathbf{vv} + {P}_{\rm g} \mathbf{I}  \right]  
 =  \bm{f}_{\rm rad} + \bm{f}_{\rm grav} + \bm{f}_{\rm vis},
\label{eq:momentum}
\end{eqnarray}
\begin{eqnarray}
\displaystyle \frac{\partial e}{\partial t}
+ \bm{\nabla} \cdot \left[ \left( e+ P_{\rm g}  \right) \bm{v}  \right]
 = - \rho {\cal L} + \bm{ v} \cdot \bm{f}_{\rm rad} + \bm{ v} \cdot \bm{f}_{\rm grav} + W_{\rm vis},
 \label{eq:energy}
\end{eqnarray}
where the total energy density is $e=P_{\rm g}/(\gamma-1)+\rho v^2/2$.
$\gamma$ denotes the specific heat ratio obtained adiabatically, i.e., $\gamma =5/3$. 
${\cal L}$ denotes the net heating/cooling rate per unit mass.
We consider the gravitational force, i.e., $\bm{f}_{\rm grav} =  - \bm{e}_r \rho G M_{\rm SMBH}/r^2 $, where $G$ denotes the gravitational constant and $r=\sqrt{R^2+z^2}$ denotes the distance from the center of SMBH with $M_{\rm SMBH}= 10^7 M_{\odot}$ and $\bm{e}_{r}$ denotes the unit vector indicating the direction away from the center.
The total density ($\rho$) is the sum of gas and dust densities, $\rho_{\text{g}}+\rho_{\text{d}}$, and $\bm{v}$ denotes the barycentric velocity, i.e., $(\rho_{\text{g}} \bm{v}_{\text{g}} + \rho_{\text{d}} \bm{v}_{\text{d}} ) / \rho$, where indexes d and g denote dust and gas, respectively.

We assume the kinematic viscosity with parameter $\alpha$ as $\nu_{\rm vis}= \alpha c_s^2/\Omega_{\rm K}$ ($\alpha$-disk model; \citealt{1973A&A....24..337S}).
The viscous force in Equation \ref{eq:momentum} and viscous heating in Equation \ref{eq:energy} are described in \citet{2005ApJ...628..368O} as

\begin{equation}
  \bm{f}_{\rm vis} = \frac{\bm{e}_{\varphi}}{R^2} \frac{\partial}{\partial R} \left[ R^2 \alpha P_{\rm g} \frac{R^2}{v_{\varphi}} \frac{\partial}{ \partial R } \left( \frac{v_{\varphi}  }{R} \right)  \right],
\end{equation}
\begin{equation}
  W_{\rm vis}= \alpha P_{\rm g} \frac{R}{v_{\varphi} } \left[  R \frac{\partial}{ \partial  R } \left( \frac{v_{\varphi}}{R} \right)  \right]^2,
\end{equation}
~\\
where $\bm{e}_{\varphi}$ denotes the unit vector indicating the azimuthal direction. 
We assume that the gas supply to the nucleus is achieved using the thin disk, 
and the viscosity parameter is set to 

\begin{equation}
\alpha=
\left\{\begin{array}{ll}
0.1 \quad &  n > 10^{3} \text{ cm}^{-3} ~\&~ T_{\textrm{g}}< 10^3 \text{ K} \\ 
0.0 \quad & {\rm otherwise}.
\end{array}\right.
\end{equation}

\subsection{Luminosity in AGN radiation sources}  \label{subsec:22}

\begin{figure}[h!]
\epsscale{1.0}
\plotone{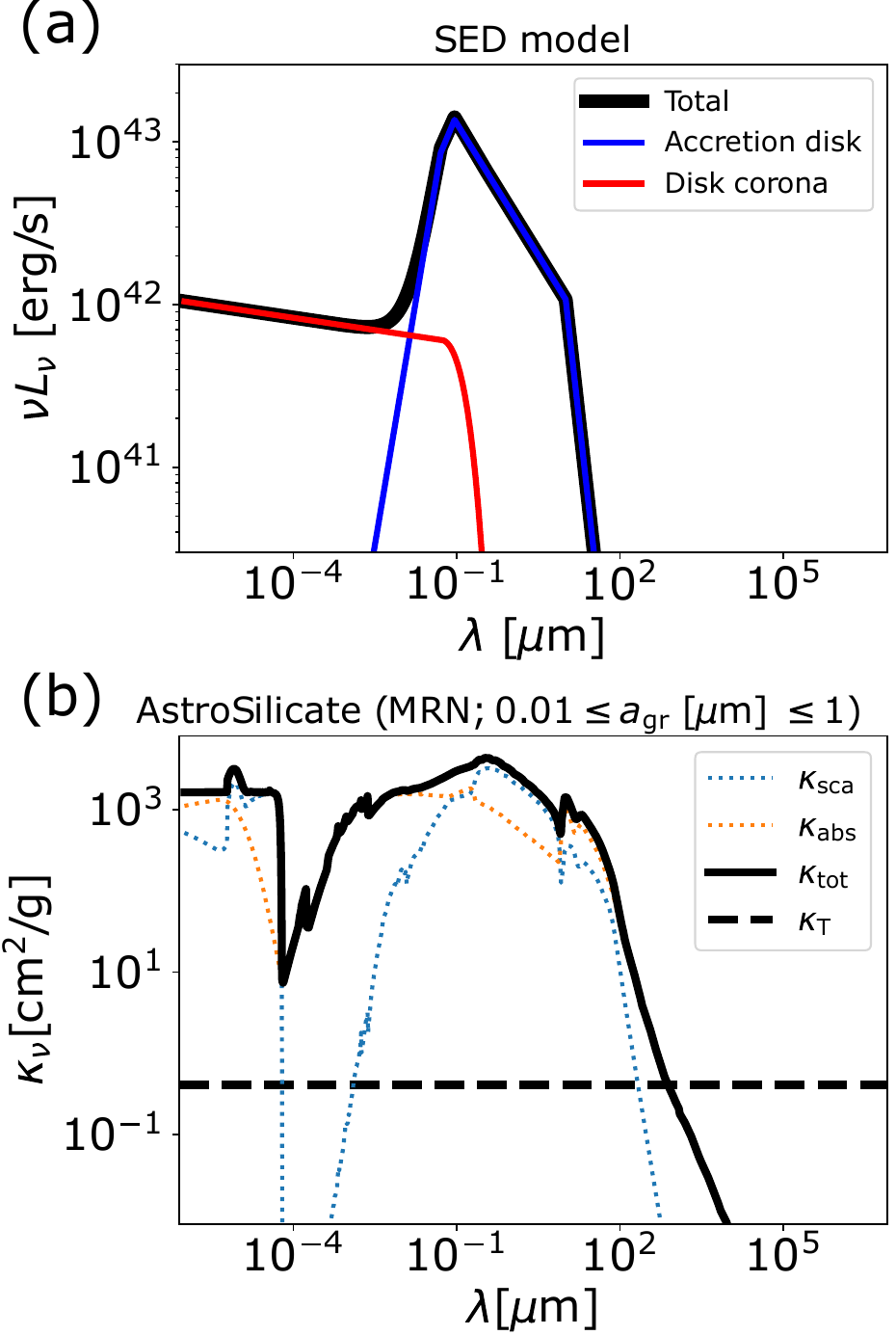}
\caption{
(a) SED model in the bolometric luminosity $L_{\text{bol}}=10^{44}$ erg s$^{-1}$. 
The accretion disk and its corona components are shown in blue and red, respectively.
(b) Total dust opacity (black solid line). $\kappa_{\rm sca}$ and $\kappa_{\rm abs}$
denote scattering and absorption opacities, respectively.
The black dashed line denotes the Thomson opacity.
\label{fig:21}}
\end{figure}

We modeled steady continuum radiation sources assuming the SEDs of accretion disk luminosity ($L_{\nu}^{\text {AD}}$) and its corona ($L_{\nu}^{\text {corona}}$) around the central SMBH.
We assumed that luminosity is proportional to the broken power law with respect to wavelength (\citealt{2016MNRAS.460..980N}), which has the accretion disk component \citep{2005A&A...437..861S,2011MNRAS.415..741S} corresponding to the standard disk \citep{1973A&A....24..337S} and the corona component $L_{\nu}^{\text {corona}} \propto \lambda^{p-1}$, where the photon index is $p=1.95$ \citep{2013MNRAS.433.2485B}.
We adopted the radiative flux ratio between 2500 \AA and 2 keV, i.e., $\alpha_{OX}=1.384$  \citep{2021ApJ...910..103L}, and the bolometric luminosity $L_{\text{bol}}=10^{44}$ erg s$^{-1}$, which is integrated over the wavelength, i.e., $L_{\text{bol}} = \int_{-\infty}^{\infty} \left( L_{\nu}^{\text{AD}}+ L_{\nu}^{\text{corona}} \right) d \nu$.
Figure \ref{fig:21}a. shows the SED used in this study.

\subsection{radiation force and radiative cooling/heating}  \label{subsec:23}
The radiation force $\bm{f}_{\text{rad}}$ in Equation (\ref{eq:momentum}) is determined based on the radiation flux $F_{\nu} $ or the amount of radiation energy absorbed by the medium in small volumes (see also Section \ref{subsec:25}),

\begin{equation}
\bm{f}_{\text{rad}} = \int_{\nu_{\text{min}}}^{\nu_{\text{max}}} \frac{ \left( \rho_{\rm d} \kappa_{{\rm d}, \nu} + \rho_{\rm g} \kappa_{\rm T}  \right)  }{c}  F_{\nu} \bm{e}_{r}  d\nu,
\label{eq:radforce}
\end{equation}
~\\
where $\nu_{\text{min}} = 10^{10} \text{ Hz}$ and $\nu_{\text{max}} = 10^{20} \text{ Hz}$.
Therefore, the radiation flux is exponentially extinctive with the optical depth $ \tau_{\nu}$,

\begin{equation}
F_{\nu}= \frac{ 1 }{4 \pi r^2} \left( L_{\nu}^{\text{AD}} q (\theta) + L_{\nu}^{\text{corona}} \right)  e^{-  \tau_{\nu} },
\label{eq:radflux}
\end{equation}
~\\
and

\begin{equation}
\tau_{\nu} = \int_{0}^r \left\{ \rho_{\rm d} \left(r^{\prime} \right) \kappa_{{\rm d}, \nu} + \rho_{\rm g} \left(r^{\prime} \right) \kappa_{\rm T} \right\} dr^{\prime},
\label{eq:opticaldepth}
\end{equation}
~\\
where the Thomson opacity is explained from electron scattering, i.e., $\kappa_{\rm T}=0.41 \text{g cm}^{-2}$.
The radiation flux for the accretion disk is anisotropic given by $q(\theta) = \sin \theta (1+ 2 \sin \theta)$ explained by limb darkening and the change in the projected surface area (\citealt{1987MNRAS.225...55N}).  
The angle $\theta$ from the equatorial plane is defined in the range of 0--$90$ deg.

We obtain dust opacity $\kappa_{\text{d}, \nu}$ based on the weighted average of dust grain sizes $a$.

\begin{equation}
\kappa_{{\rm d}, \nu}= 
\int_{a_{\text{min}}}^{a_{\text{max}}} 
\frac{4 \pi}{3} a^3 \rho_{\rm grain} n_{\rm d }(a) \kappa_{{\rm d}, \nu} (a)  da /  \rho_{\rm d},
\label{eq:opacity}
\end{equation}
~\\
where the total dust density is defined as

\begin{equation}
\rho_{\rm d}= 
\int_{a_{\text{min}}}^{a_{\text{max}}} 
\frac{4 \pi}{3} a^3 \rho_{\rm grain} n_{\rm d}(a) da.
\end{equation}
~\\
The opacity model for each grain size is based on 
Astronomical Silicate \citep{2003ApJ...598.1017D,2003ApJ...594..347D}, which provides the optical properties of  spherical carbonaceous and amorphous silicate grains, excluding polycyclic aromatic hydrocarbon molecules (PAHs), with the grain density of $\rho_{\rm grain}=3.3$ g cm$^{-3}$. 
This model includes absorption ($\kappa_{\text{abs}}$) and scattering ($\kappa_{\text{sca}}$).
Opacities are considered along the direction of radiation, i.e., $\kappa_{{\rm d}, \nu}= \kappa_{\text{tot}}= \kappa_{\text{abs}}+ \kappa_{\text{sca}}$ for Equations (\ref{eq:radforce}) and (\ref{eq:opacity}).
Here, we assume the size range of $ a_{\text{min}}=0.01 \mu \text{m}$ to $a_{\text{max}}= 1.00 \mu \text{m}$ with distribution, $dn_{\rm d} (a) / da = A n_{\rm g} a^{-\beta}$, where $\beta$ is $3.5$ \citep{1977ApJ...217..425M} and $A$ is determined by the dust-to-gas mass ratio of 0.01. 
The size range follows the assumption of reproducing the observations in \citet{2011MNRAS.415..741S}.
The opacities are shown in Figure \ref{fig:21}b.

The radiative energy received by dust and gas becomes the work done by the radiation forces ($\bm{v\cdot f}_{\text{rad}}$) and radiative heating ($\rho {\cal L}$), as expressed in Equation \ref{eq:energy}.
We consider Coulomb heating for the cold gas and Compton heating and photoionized heating for the warm and hot gas \citep{2012ApJ...758...66W}.
In particular, the radiative heating is determined by evaluating the ionization parameter $\xi= \int_{E_{\text{min}}}^{E_{\text{max}}} 4 \pi \left(n h \right)^{-1} F_{\nu}    d \left( h\nu \right)$, where $h$ is the Planck constant and the integral range is defined as $E_{\text{min}}=13.6$ eV to $E_{\text{max}}=13.6$ keV \citep[see,][]{2015ARA&A..53..365N}.
We adopted the cooling model with the metallicity of solar abundance from \cite{2005A&A...436..397M} and \cite{2009ApJ...702...63W}.
We assume the time evolution of chemistry and metallicity is ignored.

\subsection{Dust sublimation and sputtering}  \label{subsec:24}

We consider dust destruction based on the interaction with thermal gas (thermal sputtering) and AGN radiation (sublimation).
The time scale of sputtering is estimated by the approximate formula of \citet{1995ApJ...448...84T} \cite[see also, ][]{1979ApJ...231...77D, 2016MNRAS.460..980N},

\begin{eqnarray}
t_{\rm sp} &=& 5.5 \text{~yr~}  \left( \frac{n_{\text{g}}}{10^3 ~{\rm cm}^{-3} } \right)^{-1} \left( \frac{a}{0.01 ~ \mu\text{m} } \right) \\
& & \times \left[  \left( \frac{2\times 10^6 ~{\textrm{K}}}{T_{\text{g}}} \right)^{2.5} +1 \right].
\label{eq:sputtering}
\end{eqnarray}
~\\
We assume that the dusty gas immediately becomes dust-free gas if $t_{\rm sp}$ in a particular grid cell is shorter than the dynamical time scale, $t_{\text{dyn}} \sim 4.7$ yr $\left( r / 0.01 \text{ pc}  \right)^{3/2}  \left(  M_{\rm SMBH} / 10^7 M_{\odot} \right)^{-1/2}$. 
Moreover, we consider dust sublimation; dust grains are sublimated when the dust temperature exceeds $1500$ K.
The dust temperature is calculated from the local thermal equilibrium,

\begin{equation}
\begin{split}
\int \frac{L_{\nu} e^{-\tau_{\nu}} }{4 \pi r^2}  \kappa_{\text{d}, \nu}(a) \frac{4 \pi}{3} a^3 \rho_{\text{grain}} n(a) d\nu da  -  \\
\int 4 \pi  B_{\nu} (T_{\rm d})  \kappa_{\text{d}, \nu}(a) \frac{4 \pi}{3} a^3 \rho_{\text{grain}} n(a) d\nu da =0.
\end{split}
\label{eq:sublimation}
\end{equation}
~\\
The first term denotes the energy gained by the dust from AGN radiation and the second term denotes the energy lost by the blackbody radiation from the dust.

\subsection{Simulation setup} \label{subsec:25}
We use the public magnetohydrodynamics code CANS+ \citep{2019PASJ...71...83M}
\footnote{\url{https://github.com/chiba-aplab/cansplus}} 
with the additional module to evaluate the radiation force and radiative heating/cooling, without the magnetic field. 
Equations (\ref{eq:mass})--(\ref{eq:energy}) are solved for the axisymmetric cylindrical coordinate in the $R$-$z$ plane.
Moreover, we solve Equation (\ref{eq:momentum}) for the azimuthal direction as the conservation equation of the angular momentum.

From Equation (\ref{eq:radforce}), the radiation force generated owing to the irradiation flux of gas and dust (Equation (\ref{eq:radflux})) is measured by the following approximation: $\bm{f}_{\text{rad}} \simeq \int \nabla \cdot F_{\nu} \bm{e}_{r} d \nu$  \citep[e.g., ][]{2006ApJS..162..281W,2014ApJ...797....4K, 2016MNRAS.460..980N, 2017ApJ...843...58C}.
This formula means that the total amount of radiation energy in a cell can be determined by the radiation flux of the cell surface.
The radiation force can be estimated without considering optical thickness in the cell.

The optical depth is calculated by the long-characteristics method, which traces rays from the center to target cells.
However, cells in which dust destruction occurs are evaluated using Thomson opacity instead of Equation (\ref{eq:opticaldepth}), i.e., $\tau_{\nu} = \int_{r-\Delta r /2}^{r+ \Delta r/2}  \rho_{\rm g} \left(r^{\prime} \right) \kappa_{\rm T} dr^{\prime}$. 
Furthermore, we assume that, under these conditions, dust is instantaneously destroyed/solidified.

\begin{figure*}[ht!]
\epsscale{1.15}
\plotone{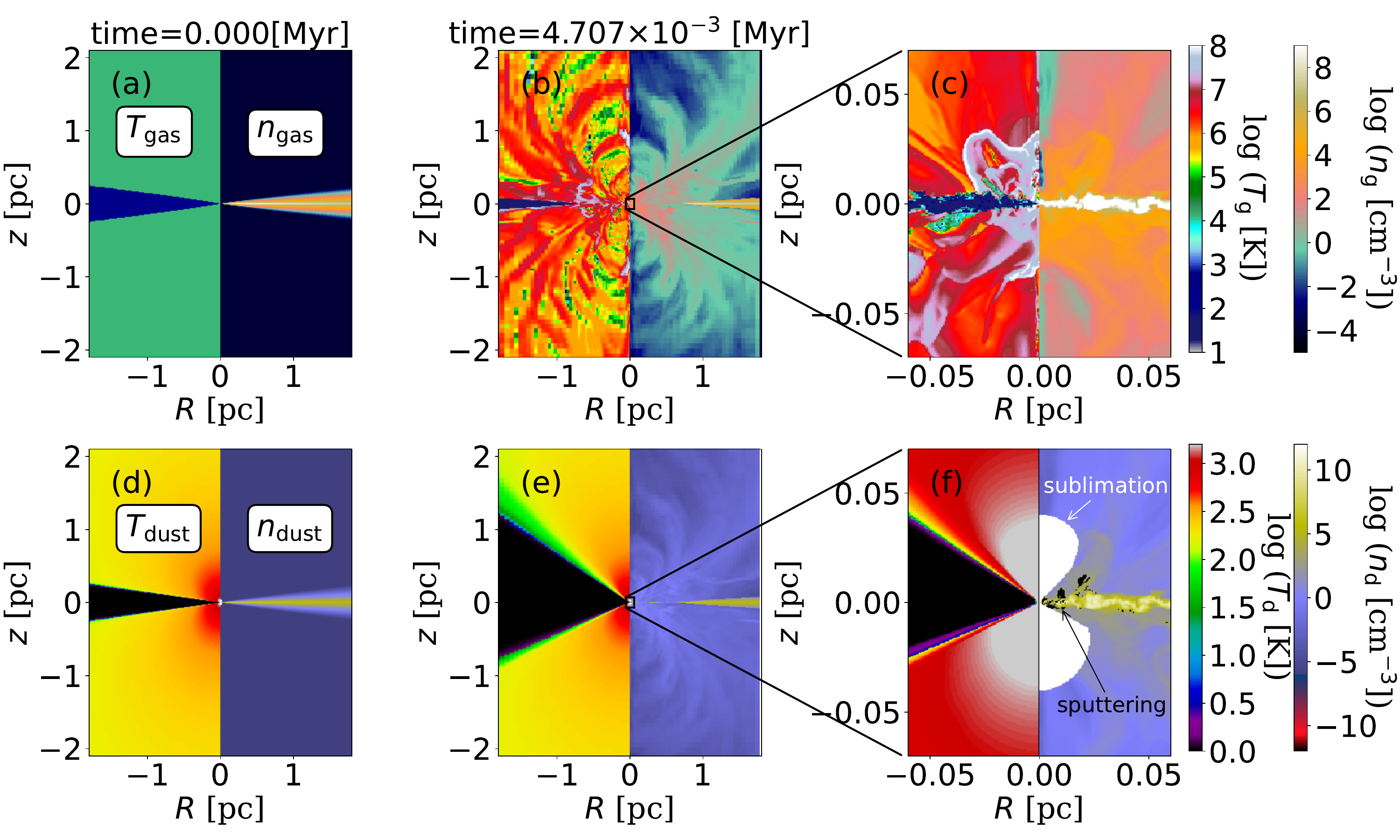}
\caption{
Evolution of the multiscale dusty outflow at t = 0 yr (initial condition); 
$t=4.71 \times 10^{3}$ yr.
In each panel, the temperature (left side) and density (right side) of the gas 
are shown.  
The panels in the top row are related to gas, and those in the bottom row are related to dust.
Panels (c) and (f) are zoomed into the sub-parsec scale of panels (b) and (e). 
In the dust number density distribution of panel (f), the dust destruction effect caused by sublimation is denoted by white, and that caused by sputtering is denoted by black. 
\label{fig:41}}
\end{figure*}

The numerical time step $\Delta t_{\textrm{num}}$ is determined by
$
\Delta t_{\textrm{num}}= C \min \left( \Delta t_{\textrm{hydro}}, \Delta t_{\textrm{rad}},  \Delta t_{\textrm{vis}} \right),
$
where the timescales are hydrodynamics $\Delta t_{\textrm{hydro}}$, radiation force $\Delta t_{\textrm{rad}} = \rho |\bm{v}| / |\bm{f}_{\textrm{rad}}| $, and $\alpha$-viscosity $\Delta t_{\textrm{vis}}= \rho v_{R} / |\bm{f}_{\textrm{vis}}| $.
In this study, we assume $C=0.2$, i.e.,  
the number of computational cells in each direction set $(N_R, N_z) = (210, 642)$.
The cell size in the uniform region is $\Delta R=\Delta z=5 \times 10^{-4}$ pc for $R\le 3.8 \times 10^{-2}$ and $|z|\le 9.6 \times 10^{-2}$ pc.
The cells at $R > 3.8 \times 10^{-2}$ pc and  $|z| > 9.6 \times 10^{-2}$ pc gradually increase to approximately 0.1 pc for the maximum simulation box $R=|z|=2$ pc.

We assume symmetries for $\rho, ~P_{\text{g}}$, and $v_z$ and anti-symmetries for $v_R$ and $v_z$ as boundary conditions for the $z$--axis. 
The outflow condition is assumed for the outer boundary.
In the central region $r \le 2 \times 10^{-3}$ pc, we adopted the absorbed boundaries as $\rho=10^{-28}$ g cm$^{-3}$, $T_{\text{g}}= 10^4$ K, and $v_{\varphi}=0$.

Initial density distribution in the Keplerian dusty-gas disk is given by

\begin{equation}
\rho(R,z)= \rho_0 \left( \frac{R}{R_0} \right)^{-p_R} \exp \left( -\frac{|z|}{H} \right) ,
\end{equation}
~\\
where the midplane density at $R_0=0.01$ pc is $\rho_0=10^{-10}$ g cm$^3$ considering the radial power $p_R=3$ (\citealt{Kawaguchi2003}) and the disk scale height is $H(R)=5\times 10^{-3} R$. 
The initial disk mass of gas and dust for $10^{-3}$ pc$ <R< 2$ pc have the  $M_{\text{g}} \simeq 1 \times 10^6 M_{\odot}$ and $M_{\text{d}} \simeq 1 \times 10^4 M_{\odot}$. 
The initial gas temperature in the disk is constant as $T_{\textrm{g}}=100$ K.
For the grid cell with $\rho<10^{-28}$ g cm$^{-3}$, we set floor values with $\rho=10^{-28}$ g cm$^{-3}$, $T_{\text{g}}=10^{4}$ K, and $\bm{v}=0$.

\begin{figure*}[h!]
\epsscale{1.15}
\plotone{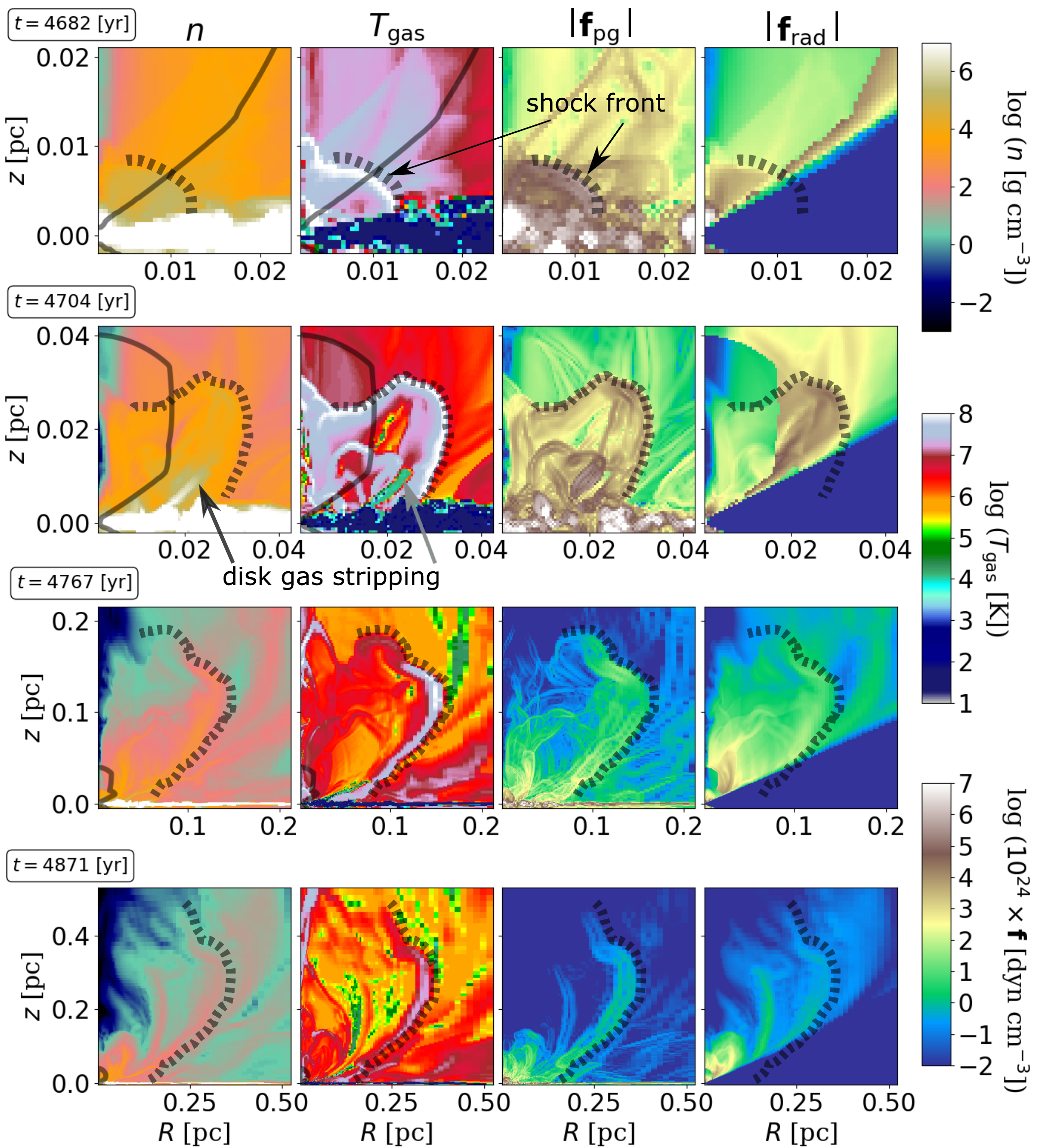}
\caption{
The time evolution of a shell-like structure associated with the multiphase gas blowout from the $10^{-2}$ pc scale (top) to the $10^{-1}$ scale (bottom).
In each snapshot from left to right, the color contours denote gas density, temperature, gas-pressure force, and radiation force, respectively.  
\label{fig:44}}
\end{figure*}

\begin{figure}[h!]
\epsscale{1.2}
\plotone{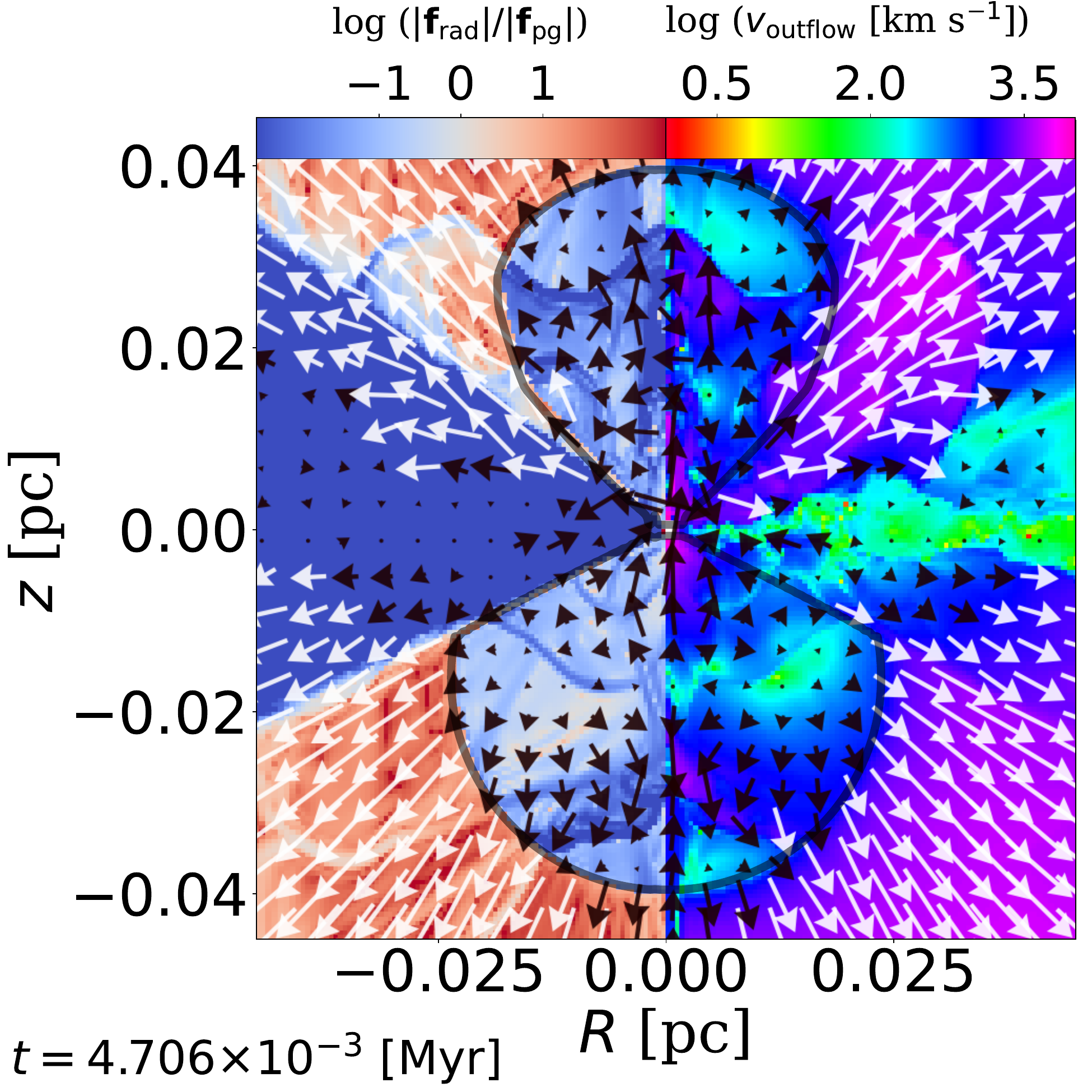}
\caption{
Left: the force ratio of radiation over gas pressure. 
Right: the magnitude of the poloidal velocity. 
The gray contour denotes the dust sublimation radius.
The arrows denote the velocity vectors in the $R$-$z$ plane; the white color represents the velocity that exceeds the escape velocity, 
and the black arrows represent the velocity that is less than the escape velocity.   
\label{fig:43}}
\end{figure}

\section{Results} \label{sec:results}
\subsection{Multiscale and multiphase outflows} \label{subsec:41}

We first overview the time evolution of gas distribution in the central 2 pc of the model considering the bolometric luminosity of $10^{44}$ erg s$^{-1}$ (Figure \ref{fig:41}). 
It shows that multiphase outflows are driven by the radiation force at $r < 0.1$ pc. 
The bottom three panels of Figure \ref{fig:41} show that dusty gas is distributed in the disk; the dust is associated with shell-like outflows, which propagate up to scales of a few parsecs.
The dusty outflows are launched from the surface of the dense dusty disk inner 0.01 pc (Figures \ref{fig:41}c and f). 
The dust temperature near the disk plane is low ($T_{\text{d}} \le 10$ K; Figures \ref{fig:41}d--f). 
This is because of the nonspherical UV radiation field (Equation \ref{eq:radflux}) and the attenuation caused by the dense gas.
As a result, the dust sublimation region is not spherical, as is evident in the white region of Figure \ref{fig:41}f, where the dust temperature exceeds the limit (1500 K).
Most dusty outflows are not fast enough to escape from the central parsec region and thus fall back to the disk, eventually infall to the central region through the disk. 
This circulation mechanism of the dusty gas is similar to the ``radiation-driven fountain”, which was proposed as a mechanism to form a geometrically thick torus-like structure on a 10 parsec scale \citep{2012ApJ...758...66W,2015ApJ...812...82W,2016ApJ...828L..19W}.

Let us see the formation process of the nonspherical multi-shell outflows in detail. 
Figure \ref{fig:44} shows the generation and propagation of a shell with regard to density, temperature, gas-pressure force, and radiation force distributions. 
The time scale of the formation and propagation of one shell is roughly $\sim 100$ yr, which is comparable to a rotational timescale ($\sim 150$ yr) at $r \sim 0.1$ pc.
Around the dust sublimation scale ($r \lesssim 0.04$ pc), the gas density and temperature are steepened by a shock with Mach number $\sim 2$, where the forces of the gas pressure ($|\bm{f}_{\text{pg}}| =\sqrt{|\bm{\nabla} P_{\text{g}}|^2}$) and radiation ($|\bm{f}_{\text{rad}}|$ of Eq.\ref{eq:radforce}) contribute to the acceleration of
outflows; these are shown by the dashed lines in Figure \ref{fig:44}. 
Beyond the dust sublimation radius, the expansion of the shock front is accelerated by the radiation force of the dust, which is greater than the gas pressure (Figure \ref{fig:43}). 
Furthermore, the effect of the anisotropic radiation transforms one shell from a spherical shape to an hourglass shape, thereby resulting in a lotus-flower-like structure with multiple outflowing shells.

According to the dust sublimation scale, the outflow associated with the expanding shells is driven by the radiation force and radiative heating.
The left-half panel of Figure \ref{fig:43} shows the force balance between radiation ($|\bm{f}_{\text{rad}}|$) and gas pressure ($|\bm{f}_{\text{pg}}|$). 
In the region inside the sublimation radius (gray solid line), the gas pressure dominates the radiation force, and the gas outflow velocity is less than the escape velocity represented by the black arrows. 
This is in contrast to the outer region, where the gas velocity exceeds the escape velocity, represented by the white arrows around the sub-parsec scale.

Notes that the sublimation radius is not symmetric above and below the disk plane, even though the radiation field is symmetric in the equatorial plane.
This asymmetric feature is always observed with the outflowing gas density and velocity field 
\footnote{
See Figures \ref{fig:41} and \ref{fig:44}, lotus-flower-like gas structure propagate independently irrespective of the disk top and bottom.
}
in the gas-pressure-dominant region of Figure \ref{fig:43}.
Symmetry is broken by the nonlinear interaction between the radiation field and gas.

\begin{figure}[h!]
\epsscale{1.2}
\plotone{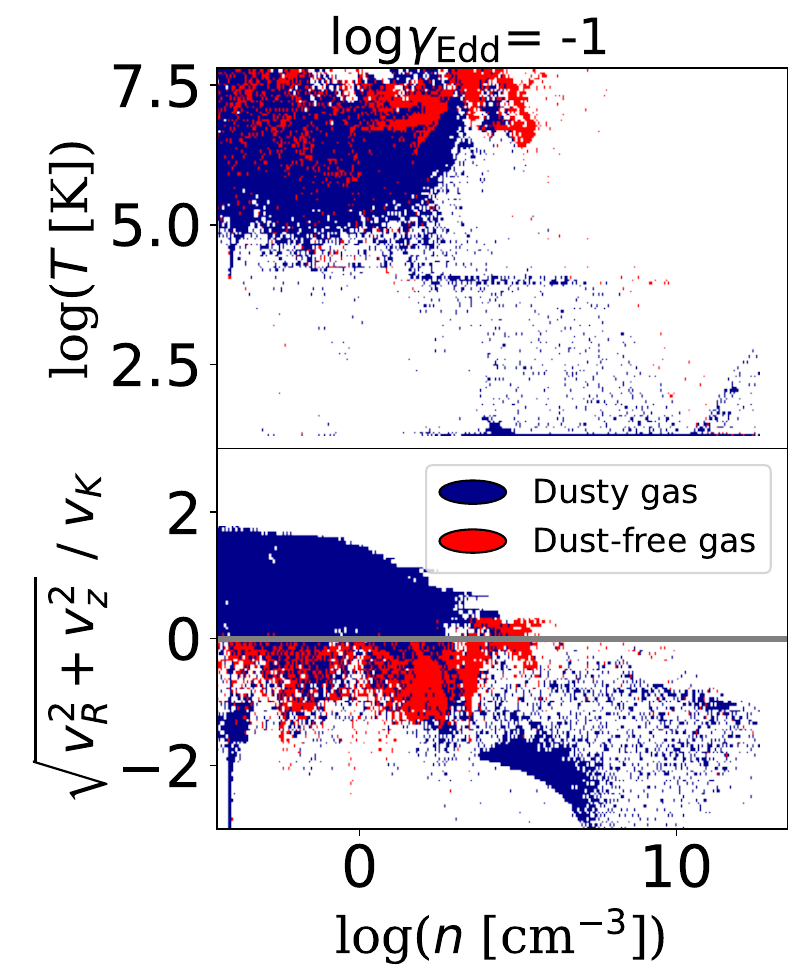}
\caption{
Phase diagram of the number density over gas temperature (top) and poloidal velocity normalized by the Keplerian rotation (bottom).
The gray line in the bottom panel denotes the criterion of the escape velocity.
The blue and red colors denote dusty and dust-free gases.
These distributions are at $t=4.71 \times 10^{3}$ yr.
} \label{fig:42}
\end{figure}

We found that the gas in the sub-parsec scale exhibits a wide range of temperature and density (Figure \ref{fig:42}).
The dust-free gas coexists with the dusty gas on the density--temperature or density--velocity plane; 
however, the hot gas ($T_{\rm g} \gtrsim 10^6$ K) with  $n \lesssim 10^6$ cm$^{-3}$ tends to be dust-free.
The dust-free gas is mostly caused by the dust sublimation due to the intense radiation field, not by the sputtering effect \footnote{
At $r=0.01$ pc,  
the spattering time scale is shorter than the dynamical time scale, i.e., 
$t_{\text{sp}}/ t_{\text{dyn}} \lesssim 1$ (eq. \ref{eq:sputtering}). 
The lower limit of number density at which sputtering occurs is $n \sim 10^3 \text{ cm}^{-3}$  for $T_{\text{g}} \geq 10^6 $ K.
At low temperatures, the density increases by a factor of $(T_{\text{g}}/  10^6 \text{K})^{-2.5}$.
}; 
the spatial distributions of sublimation and sputtering are shown in Figure \ref{fig:41}f.

Figure \ref{fig:42}(bottom) shows that the dusty gas with $n_{\text{g}}>10^6$ cm$^{-3}$ is gravitationally bound ($\sqrt{v_R^2+ v_z^2} < v_K$),
and the low-density dusty gas is gravitationally unbound.
The high and low-velocity gases spatially correspond to the gas in the radiation-dominated and gas-pressure-dominated regions shown in Figure \ref{fig:43}, respectively. 
In contrast, the velocity of the dust-free gas is less than the Keplerian velocity.

The complicated multiphase structure is a natural consequence of the interaction between the non-spherical central radiation field and the nonuniform dusty/dust-free gases. 
Moreover, shocks propagating outward peel off the cold dense gas on the disk surface (Figure \ref{fig:44}). 
This causes transition in phases from the cold dense gas ($n <10^{10}$ cm$^{-3}$) in the disk to hot diffused gas by shock-heating, 
in addition to direct heating by central radiation.

\subsection{Column density and dust sublimation radius} \label{subsec:42}
\begin{figure}[h!]
\epsscale{1.2}
\plotone{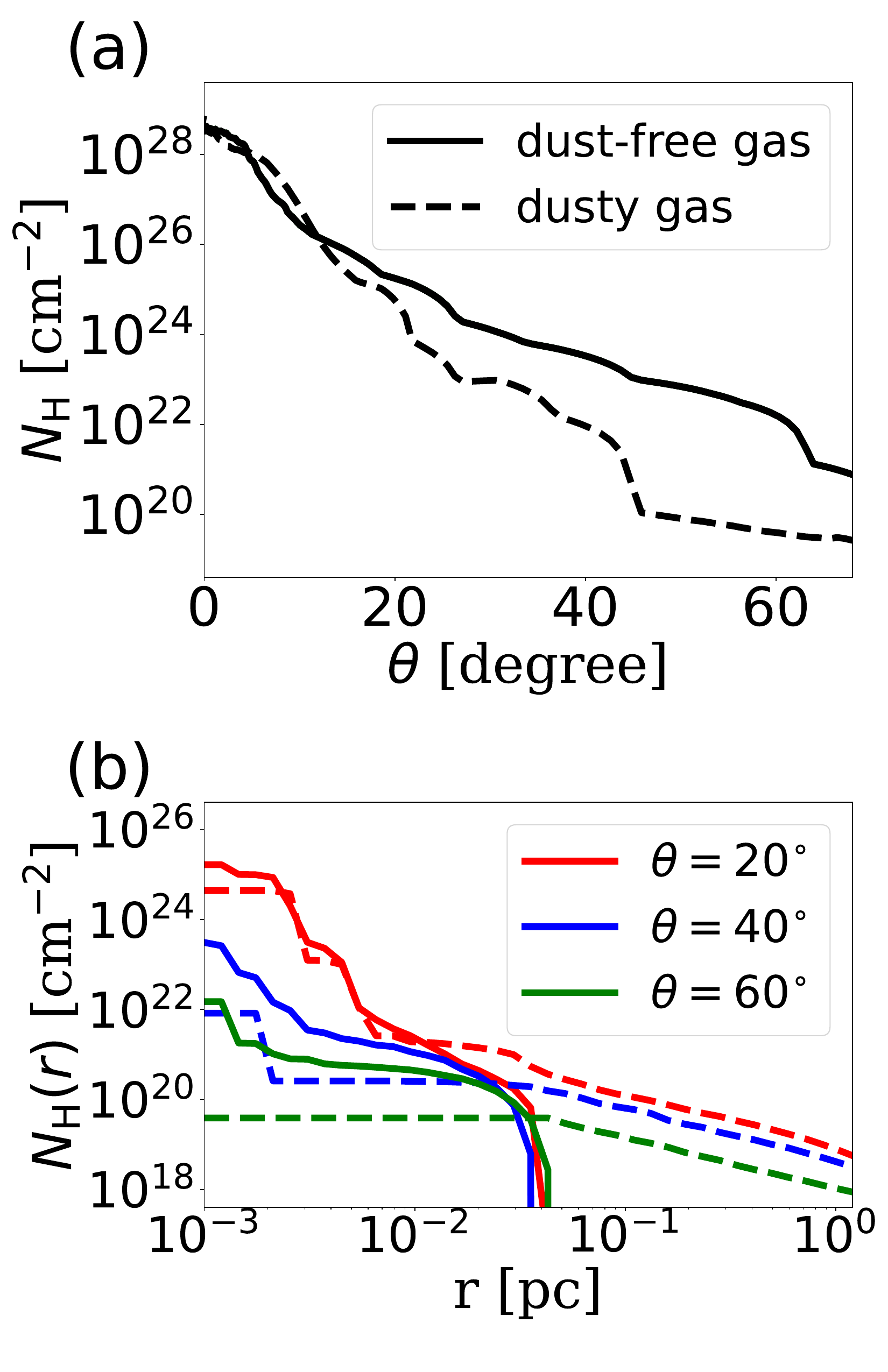}
\caption{
(a) Column density as a function of viewing angle from the equatorial plane. 
The column density is calculated using $N_{\text{H}} = \int_{r_{\text{out}}}^{0} n (l) dl$ with the starting point $r_{\text{out}}=2$ pc, where $n$ denotes the number densities of the dust-free (solid) and dusty (dashed) gases.
(b) Radial distribution of cumulative column densities $N_{\text{H}} (r) = \int_{r_{\text{out}} }^r n(l) dl$ shown in the density distribution of Figure \ref{fig:45}a.
These lines are time averaged in $4686$ yr $<t<$$4738$ yr. } 
\label{fig:46}
\end{figure}

In this section, we investigate the column density of dusty and dust-free gases and their relation to the structures of the dust sublimation region in radiation-driven outflows. 
The following discussion is based on the density field averaged over a five-rotation timescale at $R=10^{-2}$ pc to remove the time variability of the outflow (see, e.g., Figure \ref{fig:47}).

\begin{figure}[h!]
\epsscale{1.2}
\plotone{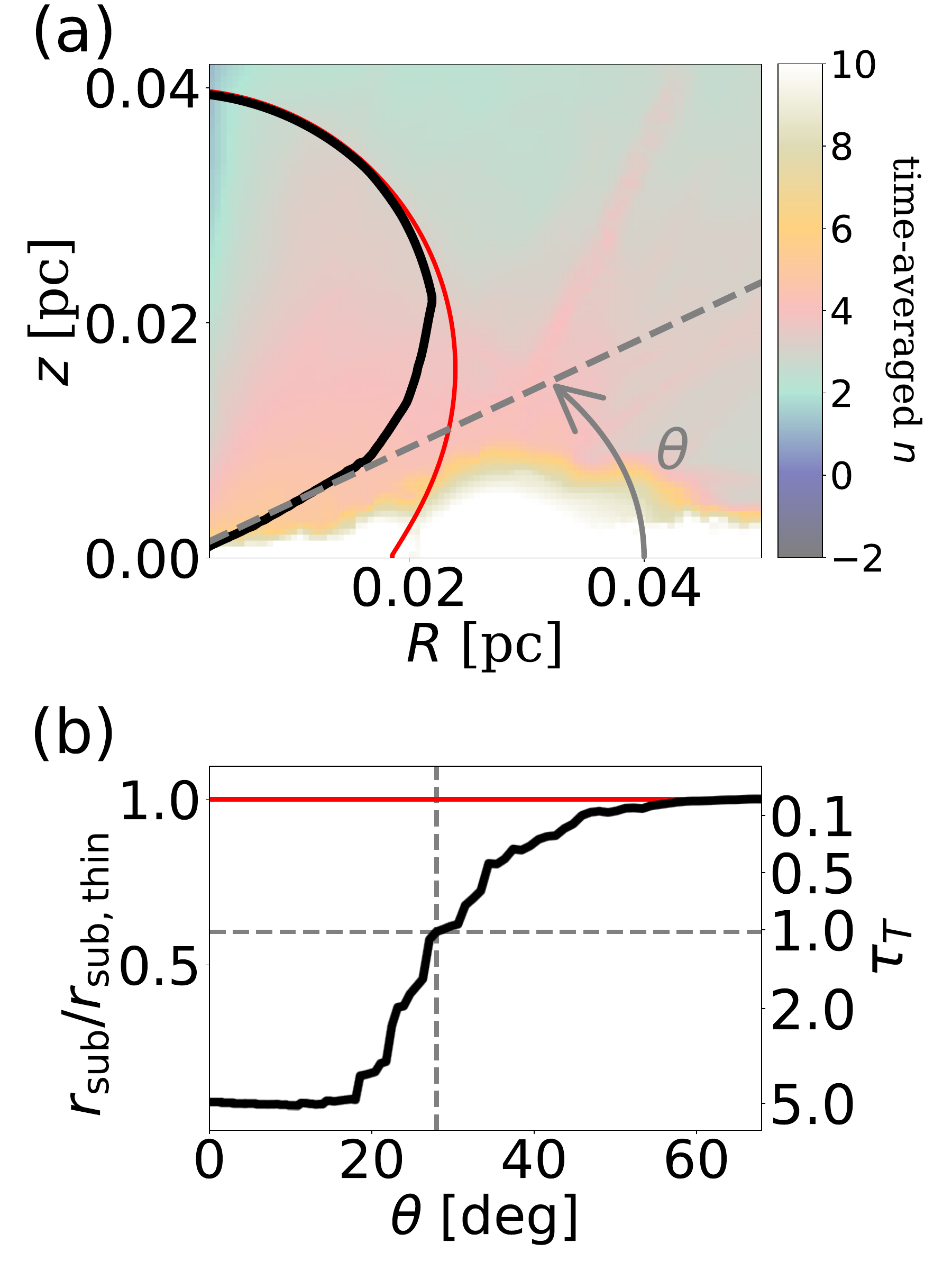}
\caption{
Dust sublimation radius as a function of angle $\theta$ at $T_{\text{d}}=1500$ K (Figs. \ref{fig:41}f and \ref{fig:43}).
(a) The black curve shows the time-averaged dust sublimation radius $r_{\text{sub}}$.
The red curve denotes the radius $r_{\text{sub, thin}}$ of the optically thin case considered for the anisotropic radiation (Eqs. \ref{eq:radflux}, \ref{eq:opticaldepth}).
The background color contour denotes the number density.
(b) The solid black curve denotes the ratio of radii normalized by the optically thin sublimation radius in panel (a) versus the specified angle. 
The gray dotted lines show the angle $\theta \sim 27$ deg where $\tau_{\text{T}} \sim 1$. 
\label{fig:45}}
\end{figure}

\begin{figure}[ht!]
\epsscale{1.2}
\plotone{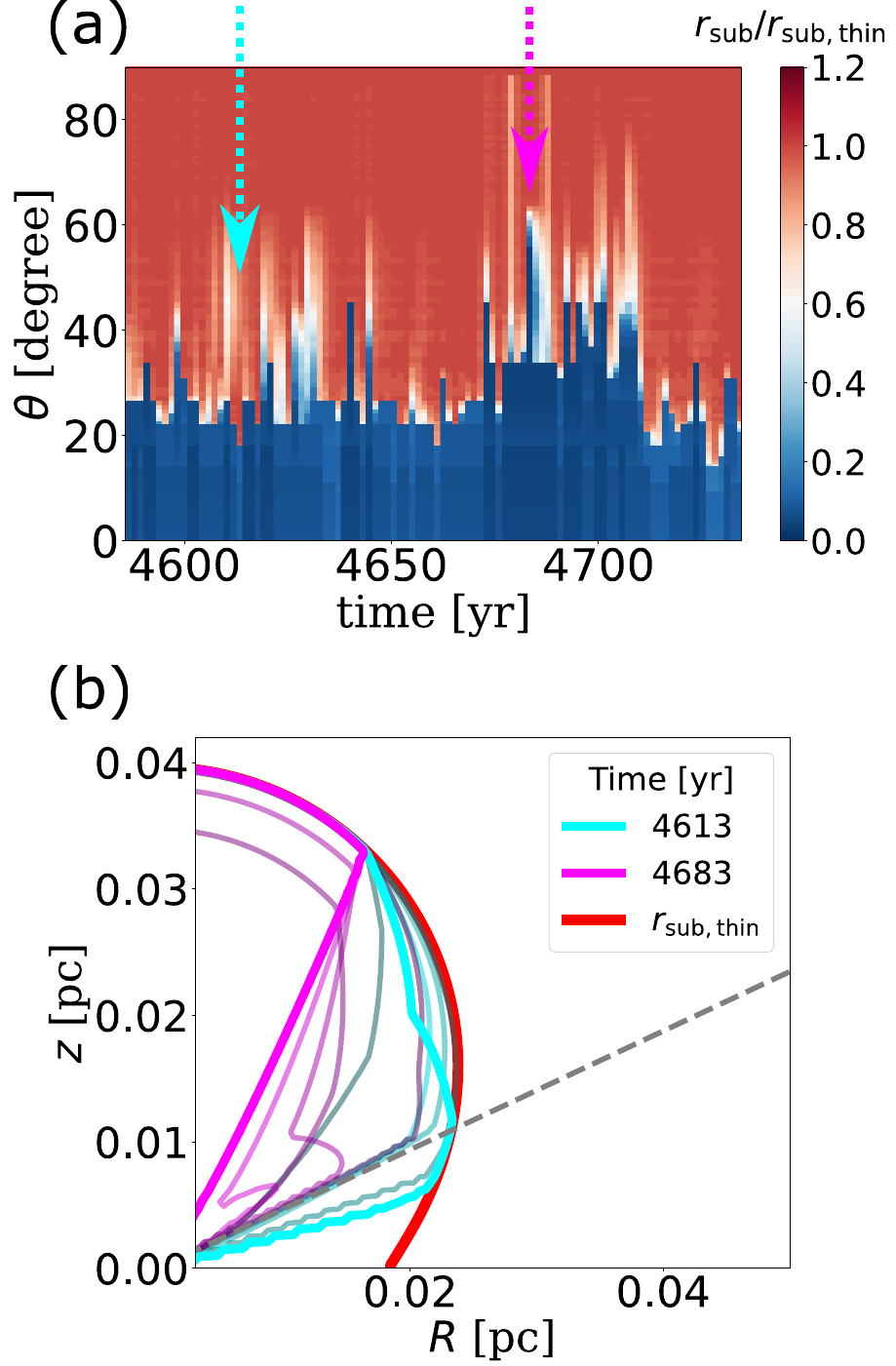}
\caption{
(a) The dust sublimation radius normalized by the optically thin sublimation radius ($r_{\rm sub, thin}$) in panel (b) shows time variability with horizontal bin $\delta t \sim 1.3$ yr.
The white region denotes the optical depth $\tau_{\text{T}} \sim 1$, and
the blue and red areas indicate optically thick and thin cases for electron scattering.
(b) The colored curves denote the snapshots of the dust sublimation radii for the two sequences. 
The cyan and magenta curves denote the dust sublimation radii of the maximum at $t=4613$ yr and the minimum at $t=4683$ yr, for which the dotted arrows in panel (a) indicate the corresponding times.
The time evolution starting from the two curves is also shown as the four snapshots, $\delta t, 2\delta t, 3\delta t,$ and $4 \delta t$.
The red curve denotes $r_{\text{sub, thin}}$ in the same as Figure \ref{fig:45}a.
The gray line denotes the angle $\theta=27$ degree, which is $\tau_{\text{T}}$ in time averaging.
}
\label{fig:47}
\end{figure}

Figure \ref{fig:46}a shows the column density distribution as a function of the angle from the disk ($\theta$) (Figure \ref{fig:45}a). 
It shows that the column density monotonically decreases from $\theta = 0$ -- $68$ deg for dusty and dust-free gases.
While the column densities of the dust-free and dusty gases are almost the same near the disk ($\theta \lesssim 10$ degree), 
these differences are evident for larger $\theta$.
Figure \ref{fig:46}b shows the column density as a function of radius for dusty and dust-free gases for three viewing angles. 
The dust-free gas appears at $r < 0.04$ pc for all the viewing angles, and the central $r \lesssim 0.01$ pc region is mostly dust-free.

The dust sublimation region in the radiation-driven outflow explained in Section \ref{subsec:41} should not be spherical, as suggested by \citet{2010ApJ...724L.183K} and \cite{2018MNRAS.474.1970B}. 
In other words, the shape of the dust sublimation region is not expressed by a single critical radius, i.e.,  $r_{\text{sub, thin}} \propto L_{\text{bol}}^{0.5}$, where a point source and optically thin material are assumed.

In Figure \ref{fig:45},  we show the difference in the sublimation radii of the optically thin case (red line) and that derived from the numerical result considering the central 0.04 pc.
The latter is always less than the optically thin case
\footnote{
The sublimation radius for the optically thin case is not zero for the disk plane because the X-ray radiation of the central source is assumed to be spherical (see also Equation (\ref{eq:radflux})).
}.
The decrease in the size of the sublimation radius is evident owing to $\theta < 27$ deg (Figure \ref{fig:45}a ).
This difference is caused by the attenuation of the dust-free gas, i.e., Thomson scattering.
In Figure \ref{fig:45}b, we show the ratio of the sublimation radii as a function of $\theta$ for the two cases; 
this ratio corresponds to the optical depth $\tau_T$, i.e., $r_{\text{sub}}/r_{\text{sub, thin} } = \exp \left( - \tau_{\text{T}} \right)$.
We found that $\tau_{\text{T}}=1$ is for $\theta = 27$ degree.
This angle corresponds to that for $N_{\text{H}} = \kappa_{\text{T}} m / \tau_{\text {T}} = 10^{24}$ cm$^{-2}$ (Figure \ref{fig:46}a).
The result implies that dust-free gas is an essential component that determines the shape of the dust sublimation region.

We found that the dust sublimation radius in the simulation shows time variation (Figure \ref{fig:47}a), where $r_{\text {sub} }/r_{\text{sub, thin}}$ is plotted as a function of time and $\theta$.
The ratio is equivalent to the optical depth for the Thomson opacity (see also conversion in the dual axis of Figure \ref{fig:45}b). 
This is also confirmed in the radii of $R$--$z$ plane (Figure \ref{fig:47}b).
Contrary to the optically thin case (red curve: $r_{\text{sub, thin}}(\theta)$), 
the sublimation radius calculated from the gas distribution inside $r = r_{\text{sub, thin}}$ is not static, and its variation depends on $\theta$. 
For $20$ deg $\lesssim \theta \lesssim 60$ deg, the sublimation radius fluctuates on a 
timescale $\lesssim$ a few years, reflecting the nonsteady outflow.

\section{Discussion}  \label{sec:discussion}

\begin{figure}[h!]
\epsscale{1.2}
\plotone{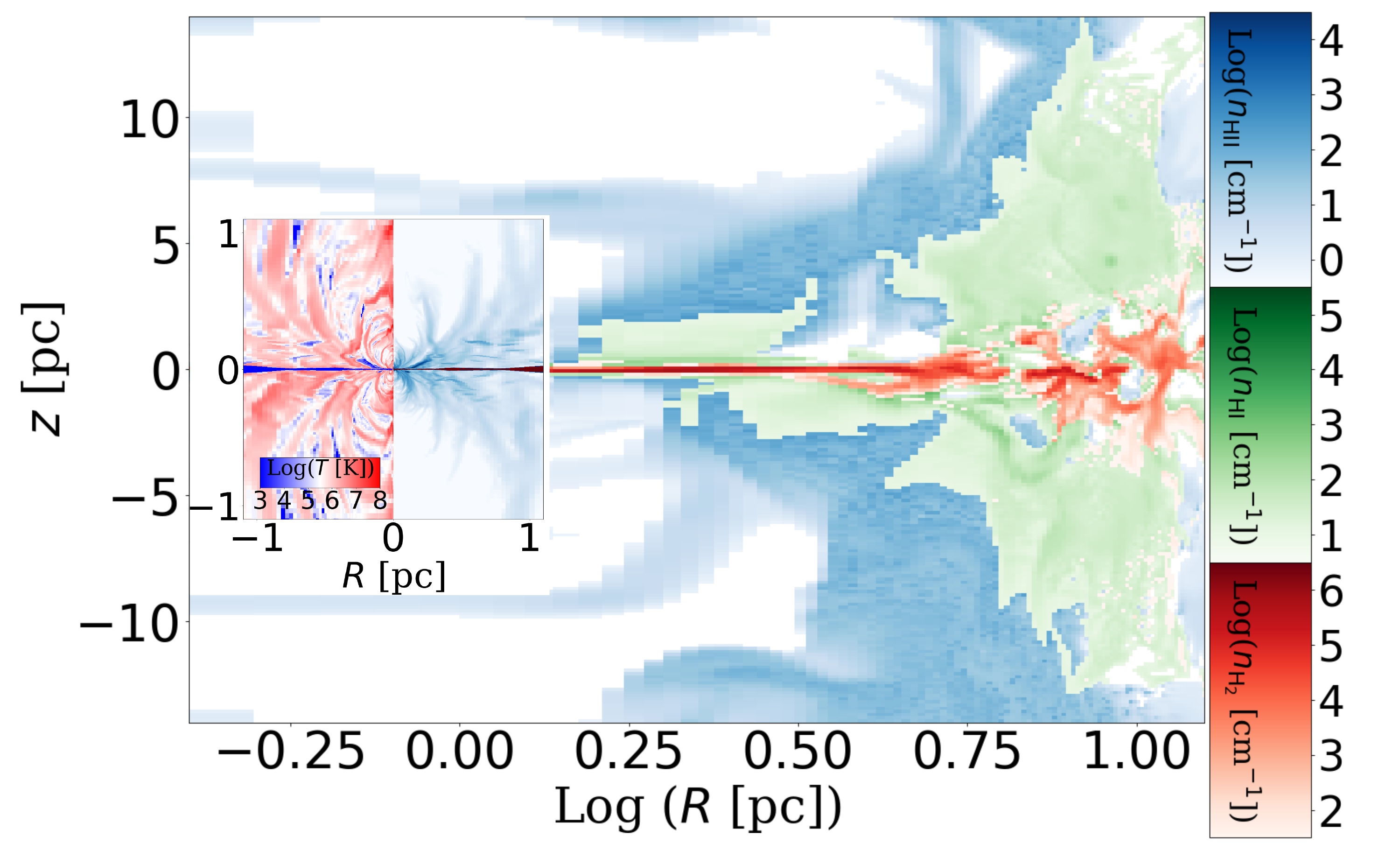}
\caption{Multiphase gas structures based on the present model (sub-parsec scale) and a
10 parsec scale \citep{2016ApJ...828L..19W}.  The molecular, atomic, and ionized gas densities are represented by red, green, and blue colors, respectively.}
\label{fig:55}
\end{figure}

The outer radius of the simulation box in our model is 2 pc. Therefore, we cannot follow fate of the outflow on a torus scale (i.e., 10 pc).
Figure \ref{fig:55} shows the multiphase gas structures based on two different scales: the present model and radiative-driven fountain model in \citet{2016ApJ...828L..19W}.
If the mass is supplied through the geometrically thin molecular dusty disk from 10 to 0.1 pc, 
ionized multi-shell outflow is formed. 
This inflow-induced outflow may propagate outward and form the ionized gas envelope (e.g., warm absorbers \citealt{2022ApJ...925...55O} and the narrow emission line region \citealt{2018ApJ...867...49W}). 
The stratified multiphase structure is actually consistent with recent high-resolution ALMA observations in the nearest type-2 Seyfert galaxy, the Circinus galaxy \citep{2023arXiv230503993I}.

In the radiation-driven fountain model \citep{2014MNRAS.445.3878S,2016ApJ...828L..19W}, the observed polar dust emission in mid-IR \citep{2014A&A...563A..82T,2016A&A...591A..47L,2022A&A...663A..35I} originated in the dusty outflow. 
However, the innermost region of the outflow (i.e., 0.1 pc) was not resolved in these simulations. 
Our model suggests that the dusty outflow originates inside the dust sublimation radius, i.e., 0.04 pc on the scale.
The high-density dust-free gas inside $r \sim 0.04$ pc (Figure \ref{fig:46}) could be the origin of Compton-thick obscuration, which has been confirmed by the X-ray observations
of nearby AGNs \citep{2015ApJ...806..127D,2019ApJ...870...31I,2021A&A...651A..58B,2022ApJ...925...55O}.
We suggest that the dust sublimation radius, which varies on the timescale below a few years, contributes to the variability observed in X-ray \citep[e.g.,][]{2020MNRAS.495.2921N,2022ApJ...927..227L} as related to IR \citep[e.g.,][]{2011A&A...527A.121K,2014ApJ...788..159K,2019ApJ...886..150M}.

Furthermore, our results imply that a hot dust ($T_{\text{d}}\sim 1000$ K) component in a central sub-parsec region can exist in the near-IR source, we expect to be observed by molecular CO absorption lines in ultra-luminous infrared galaxies (e.g.  \citealt{2018ApJ...852...83B,2022ApJ...928..184B,2021ApJ...921..141O}) and Seyfert galaxies (\citealt{2021ApJ...915...89U,2022ApJ...934...25M}).

In this paper, we investigated a case for Seyfert-like AGN with an SMBH of $10^7 M_{\odot}$. 
For more massive BHs, the outflow structures may change, as \citealt{2015ApJ...812...82W} suggested, in which the radiation-driven outflows on a scale for various parameters (e.g., the Eddington ratio and SMBH mass) are investigated. 
They found that the opening angle of the outflow tends to be larger for a larger Eddington ratio. 
This should also be the case for the sub-parsec scale outflows. 
However, we expect that the size of the dust-sublimation region may affect the launching
of the sub-parsec scale outflows.
In subsequent papers, we will investigate how outflows change depending on the Eddington ratio, SMBH mass, and SEDs.

The inflow-induced outflow presented in this paper is suitable for standard accretion disk assuming anisotropy of the central radiation field. 
The SED of the central source varies with surface density and accretion rate \citep[see, e.g.,][]{1995ApJ...438L..37A}. 
Therefore, the outflow structure may depend on the accretion disk model and AGN type.
This remains an issue for future work.

\section{Summary}  \label{sec:summary} 

We investigated the dusty outflow driven by AGN radiation fields with the bolometric luminosity $10^{44}$ erg s$^{-1}$ and SMBH mass $10^7 M_{\odot}$.
We performed radiation-hydrodynamic 2D simulations for $R = 10^{-3}$--2 pc, in which
the dust sublimation scale is resolved. 

We conclude on the following points:
\begin{itemize}
   \item[(1)]
The dusty gas in the outflow is intermittently launched from the geometrically thin disk of $r \lesssim 0.04$ pc forming ``lotus-like'' multiple shells. 
Each shell is accompanied by shock where the number density, temperature, and Mach number are $n\sim 10^{5}$ cm$^{-3}$, $T_{\text{g}}>10^7$ K, and $\sim 2$ at 0.01 pc, respectively.
Because of the anisotropic radiation in the thin accretion disk, shells propagated to a parsec scale are deformed to an hourglass-like shape.
Moreover, we observed that, within 20 degrees from the disk midplane, the outflow velocity of the dusty gas is less than the escape velocity.
The gas pressure dominates this region.
   \item[(2)]
The column density is determined by dust-free gas.
Below 27 degrees, there are optically thick dusty/dust-free gases, i.e., the column density is $N_{\text{H}} \gtrsim 10^{24}$ cm$^{-2}$.
For $N_{\text{H}} < 10^{24}$ cm$^{-2}$, the column density is a smaller fraction of dusty gas compared to dust-free gas.
The dust-free gas outflows in the central region $r<0.04$ pc, where $n \lesssim10^6$ cm$^{-3}$, $T_{\rm g}> 10^6$ K, and $v_{r} < v_{\text{esc}} $. 
   \item[(3)]
The shape of the dust sublimation radius is determined not only by the radiation flux density and its anisotropy, but also by the obscuration of the dust-free gas. 
Moreover, we found that the sublimation radius at $<60$ degrees varies with time on the scale of a few years or less owing to the nonsteady dusty outflow.   
\end{itemize}

We found that the inflow-induced dusty outflow is multiphase-stratified from the molecular cold disk to the atomic gas and formed an ionized envelope owing to the AGN radiation. 
This is consistent with the recent high-resolution ALMA observations of the nearest Seyfert galaxy, the Circinus galaxy \citep{2023arXiv230503993I}.

\acknowledgments
We thank the anonymous reference for his/her variable suggestions to improve the manuscript.
Numerical computations were carried out on Cray XC50 at the Center for Computational Astrophysics, National Astronomical Observatory of Japan.
For the parameter survey of the numerical model, this study also used computational resources of the supercomputer Fugaku provided by RIKEN through the HPCI System Research Project (Project ID: hp200234, hp210147, hp210164).
This study was supported by JSPS KAKENHI grant No. 19K03918 (N.K.), JP20K14525 (M.N.), and 21H04496 (K.W.).
Y.K. and K.W. were supported by NAOJ ALMA Scientific Research grant No. 2020-14A.


%

\vspace{5mm}


\software{CANS+\citep{2019PASJ...71...83M}
          }




\bibliography{bibtex2DRHD}

\begin{thebibliography}{}
\expandafter\ifx\csname natexlab\endcsname\relax\def\natexlab#1{#1}\fi
\providecommand{\url}[1]{\href{#1}{#1}}
\providecommand{\dodoi}[1]{doi:~\href{http://doi.org/#1}{\nolinkurl{#1}}}
\providecommand{\doeprint}[1]{\href{http://ascl.net/#1}{\nolinkurl{http://ascl.net/#1}}}
\providecommand{\doarXiv}[1]{\href{https://arxiv.org/abs/#1}{\nolinkurl{https://arxiv.org/abs/#1}}}

\bibitem[{{Abramowicz} {et~al.}(1995){Abramowicz}, {Chen}, {Kato}, {Lasota}, \&
  {Regev}}]{1995ApJ...438L..37A}
{Abramowicz}, M.~A., {Chen}, X., {Kato}, S., {Lasota}, J.-P., \& {Regev}, O.
  1995, \apjl, 438, L37, \dodoi{10.1086/187709}

\bibitem[{{Almeyda} {et~al.}(2020){Almeyda}, {Robinson}, {Richmond}, {Nikutta},
  \& {McDonough}}]{2020ApJ...891...26A}
{Almeyda}, T., {Robinson}, A., {Richmond}, M., {Nikutta}, R., \& {McDonough},
  B. 2020, \apj, 891, 26, \dodoi{10.3847/1538-4357/ab6aa1}

\bibitem[{{Almeyda} {et~al.}(2017){Almeyda}, {Robinson}, {Richmond}, {Vazquez},
  \& {Nikutta}}]{2017ApJ...843....3A}
{Almeyda}, T., {Robinson}, A., {Richmond}, M., {Vazquez}, B., \& {Nikutta}, R.
  2017, \apj, 843, 3, \dodoi{10.3847/1538-4357/aa7687}

\bibitem[{{Antonucci}(1993)}]{1993ARA&A..31..473A}
{Antonucci}, R. 1993, \araa, 31, 473,
  \dodoi{10.1146/annurev.aa.31.090193.002353}

\bibitem[{{Baba} {et~al.}(2018){Baba}, {Nakagawa}, {Isobe}, \&
  {Shirahata}}]{2018ApJ...852...83B}
{Baba}, S., {Nakagawa}, T., {Isobe}, N., \& {Shirahata}, M. 2018, \apj, 852,
  83, \dodoi{10.3847/1538-4357/aa9f25}

\bibitem[{{Baba} {et~al.}(2022){Baba}, {Imanishi}, {Izumi}, {Kawamuro},
  {Nguyen}, {Nakagawa}, {Isobe}, {Onishi}, \&
  {Matsumoto}}]{2022ApJ...928..184B}
{Baba}, S., {Imanishi}, M., {Izumi}, T., {et~al.} 2022, \apj, 928, 184,
  \dodoi{10.3847/1538-4357/ac57c2}

\bibitem[{{Barvainis}(1987)}]{1987ApJ...320..537B}
{Barvainis}, R. 1987, \apj, 320, 537, \dodoi{10.1086/165571}

\bibitem[{{Baskin} \& {Laor}(2018)}]{2018MNRAS.474.1970B}
{Baskin}, A., \& {Laor}, A. 2018, \mnras, 474, 1970,
  \dodoi{10.1093/mnras/stx2850}

\bibitem[{{Brightman} {et~al.}(2013){Brightman}, {Silverman}, {Mainieri},
  {Ueda}, {Schramm}, {Matsuoka}, {Nagao}, {Steinhardt}, {Kartaltepe},
  {Sanders}, {Treister}, {Shemmer}, {Brandt}, {Brusa}, {Comastri}, {Ho},
  {Lanzuisi}, {Lusso}, {Nandra}, {Salvato}, {Zamorani}, {Akiyama}, {Alexander},
  {Bongiorno}, {Capak}, {Civano}, {Del Moro}, {Doi}, {Elvis}, {Hasinger},
  {Laird}, {Masters}, {Mignoli}, {Ohta}, {Schawinski}, \&
  {Taniguchi}}]{2013MNRAS.433.2485B}
{Brightman}, M., {Silverman}, J.~D., {Mainieri}, V., {et~al.} 2013, \mnras,
  433, 2485, \dodoi{10.1093/mnras/stt920}

\bibitem[{{Buchner} {et~al.}(2021){Buchner}, {Brightman}, {Balokovi{\'c}},
  {Wada}, {Bauer}, \& {Nandra}}]{2021A&A...651A..58B}
{Buchner}, J., {Brightman}, M., {Balokovi{\'c}}, M., {et~al.} 2021, \aap, 651,
  A58, \dodoi{10.1051/0004-6361/201834963}

\bibitem[{{Chan} \& {Krolik}(2016)}]{2016ApJ...825...67C}
{Chan}, C.-H., \& {Krolik}, J.~H. 2016, \apj, 825, 67,
  \dodoi{10.3847/0004-637X/825/1/67}

\bibitem[{{Chan} \& {Krolik}(2017)}]{2017ApJ...843...58C}
---. 2017, \apj, 843, 58, \dodoi{10.3847/1538-4357/aa76e4}

\bibitem[{{Combes}(2021)}]{2021agnf.book.....C}
{Combes}, F. 2021, {Active Galactic Nuclei: Fueling and Feedback},
  \dodoi{10.1088/2514-3433/ac2a27}

\bibitem[{{Davies} {et~al.}(2015){Davies}, {Burtscher}, {Rosario},
  {Storchi-Bergmann}, {Contursi}, {Genzel}, {Graci{\'a}-Carpio}, {Hicks},
  {Janssen}, {Koss}, {Lin}, {Lutz}, {Maciejewski}, {M{\"u}ller-S{\'a}nchez},
  {Orban de Xivry}, {Ricci}, {Riffel}, {Riffel}, {Schartmann},
  {Schnorr-M{\"u}ller}, {Sternberg}, {Sturm}, {Tacconi}, \&
  {Veilleux}}]{2015ApJ...806..127D}
{Davies}, R.~I., {Burtscher}, L., {Rosario}, D., {et~al.} 2015, \apj, 806, 127,
  \dodoi{10.1088/0004-637X/806/1/127}

\bibitem[{{Dorodnitsyn} {et~al.}(2016){Dorodnitsyn}, {Kallman}, \&
  {Proga}}]{2016ApJ...819..115D}
{Dorodnitsyn}, A., {Kallman}, T., \& {Proga}, D. 2016, \apj, 819, 115,
  \dodoi{10.3847/0004-637X/819/2/115}

\bibitem[{{Draine}(2003)}]{2003ApJ...598.1017D}
{Draine}, B.~T. 2003, \apj, 598, 1017, \dodoi{10.1086/379118}

\bibitem[{{Draine} \& {Salpeter}(1979)}]{1979ApJ...231...77D}
{Draine}, B.~T., \& {Salpeter}, E.~E. 1979, \apj, 231, 77,
  \dodoi{10.1086/157165}

\bibitem[{{Draine} \& {Tan}(2003)}]{2003ApJ...594..347D}
{Draine}, B.~T., \& {Tan}, J.~C. 2003, \apj, 594, 347, \dodoi{10.1086/376855}

\bibitem[{{Garc{\'\i}a-Burillo} {et~al.}(2019){Garc{\'\i}a-Burillo}, {Combes},
  {Ramos Almeida}, {Usero}, {Alonso-Herrero}, {Hunt}, {Rouan}, {Aalto},
  {Querejeta}, {Viti}, {van der Werf}, {Vives-Arias}, {Fuente}, {Colina},
  {Mart{\'\i}n-Pintado}, {Henkel}, {Mart{\'\i}n}, {Krips}, {Gratadour}, {Neri},
  \& {Tacconi}}]{2019A&A...632A..61G}
{Garc{\'\i}a-Burillo}, S., {Combes}, F., {Ramos Almeida}, C., {et~al.} 2019,
  \aap, 632, A61, \dodoi{10.1051/0004-6361/201936606}

\bibitem[{{Garc{\'\i}a-Burillo} {et~al.}(2021){Garc{\'\i}a-Burillo},
  {Alonso-Herrero}, {Ramos Almeida}, {Gonz{\'a}lez-Mart{\'\i}n}, {Combes},
  {Usero}, {H{\"o}nig}, {Querejeta}, {Hicks}, {Hunt}, {Rosario}, {Davies},
  {Boorman}, {Bunker}, {Burtscher}, {Colina}, {D{\'\i}az-Santos}, {Gandhi},
  {Garc{\'\i}a-Bernete}, {Garc{\'\i}a-Lorenzo}, {Ichikawa}, {Imanishi},
  {Izumi}, {Labiano}, {Levenson}, {L{\'o}pez-Rodr{\'\i}guez}, {Packham},
  {Pereira-Santaella}, {Ricci}, {Rigopoulou}, {Rouan}, {Shimizu}, {Stalevski},
  {Wada}, \& {Williamson}}]{2021A&A...652A..98G}
{Garc{\'\i}a-Burillo}, S., {Alonso-Herrero}, A., {Ramos Almeida}, C., {et~al.}
  2021, \aap, 652, A98, \dodoi{10.1051/0004-6361/202141075}

\bibitem[{{Gravity Collaboration} {et~al.}(2020){Gravity Collaboration},
  {Dexter}, {Shangguan}, {H{\"o}nig}, {Kishimoto}, {Lutz}, {Netzer}, {Davies},
  {Sturm}, {Pfuhl}, {Amorim}, {Baub{\"o}ck}, {Brandner}, {Cl{\'e}net}, {de
  Zeeuw}, {Eckart}, {Eisenhauer}, {F{\"o}rster Schreiber}, {Gao}, {Garcia},
  {Genzel}, {Gillessen}, {Gratadour}, {Jim{\'e}nez-Rosales}, {Lacour},
  {Millour}, {Ott}, {Paumard}, {Perraut}, {Perrin}, {Peterson}, {Petrucci},
  {Prieto}, {Rouan}, {Schartmann}, {Shimizu}, {Sternberg}, {Straub},
  {Straubmeier}, {Tacconi}, {Tristram}, {Vermot}, {Waisberg}, {Widmann}, \&
  {Woillez}}]{2020A&A...635A..92G}
{Gravity Collaboration}, {Dexter}, J., {Shangguan}, J., {et~al.} 2020, \aap,
  635, A92, \dodoi{10.1051/0004-6361/201936767}

\bibitem[{{H{\"o}nig}(2019)}]{2019ApJ...884..171H}
{H{\"o}nig}, S.~F. 2019, \apj, 884, 171, \dodoi{10.3847/1538-4357/ab4591}

\bibitem[{{H{\"o}nig} {et~al.}(2018){H{\"o}nig}, {Alonso Herrero}, {Gandhi},
  {Kishimoto}, {Pott}, {Ramos Almeida}, {Surdej}, \&
  {Tristram}}]{2018ExA....46..413H}
{H{\"o}nig}, S.~F., {Alonso Herrero}, A., {Gandhi}, P., {et~al.} 2018,
  Experimental Astronomy, 46, 413, \dodoi{10.1007/s10686-018-9612-3}

\bibitem[{{H{\"o}nig} \& {Kishimoto}(2011)}]{2011A&A...534A.121H}
{H{\"o}nig}, S.~F., \& {Kishimoto}, M. 2011, \aap, 534, A121,
  \dodoi{10.1051/0004-6361/201117750}

\bibitem[{{Ichikawa} {et~al.}(2019){Ichikawa}, {Ricci}, {Ueda}, {Bauer},
  {Kawamuro}, {Koss}, {Oh}, {Rosario}, {Shimizu}, {Stalevski}, {Fuller},
  {Packham}, \& {Trakhtenbrot}}]{2019ApJ...870...31I}
{Ichikawa}, K., {Ricci}, C., {Ueda}, Y., {et~al.} 2019, \apj, 870, 31,
  \dodoi{10.3847/1538-4357/aaef8f}

\bibitem[{{Impellizzeri} {et~al.}(2019){Impellizzeri}, {Gallimore}, {Baum},
  {Elitzur}, {Davies}, {Lutz}, {Maiolino}, {Marconi}, {Nikutta}, {O'Dea}, \&
  {Sani}}]{2019ApJ...884L..28I}
{Impellizzeri}, C.~M.~V., {Gallimore}, J.~F., {Baum}, S.~A., {et~al.} 2019,
  \apjl, 884, L28, \dodoi{10.3847/2041-8213/ab3c64}

\bibitem[{{Isbell} {et~al.}(2022){Isbell}, {Meisenheimer}, {Pott}, {Stalevski},
  {Tristram}, {Sanchez-Bermudez}, {Hofmann}, {G{\'a}mez Rosas}, {Jaffe},
  {Burtscher}, {Leftley}, {Petrov}, {Lopez}, {Henning}, {Weigelt}, {Allouche},
  {Berio}, {Bettonvil}, {Cruzalebes}, {Dominik}, {Heininger}, {Hogerheijde},
  {Lagarde}, {Lehmitz}, {Matter}, {Meilland}, {Millour}, {Robbe-Dubois},
  {Schertl}, {van Boekel}, {Varga}, \& {Woillez}}]{2022A&A...663A..35I}
{Isbell}, J.~W., {Meisenheimer}, K., {Pott}, J.~U., {et~al.} 2022, \aap, 663,
  A35, \dodoi{10.1051/0004-6361/202243271}

\bibitem[{{Izumi} {et~al.}(2018){Izumi}, {Wada}, {Fukushige}, {Hamamura}, \&
  {Kohno}}]{2018ApJ...867...48I}
{Izumi}, T., {Wada}, K., {Fukushige}, R., {Hamamura}, S., \& {Kohno}, K. 2018,
  \apj, 867, 48, \dodoi{10.3847/1538-4357/aae20b}

\bibitem[{{Izumi} {et~al.}(2023){Izumi}, {Wada}, {Imanishi}, {Nakanishi},
  {Kohno}, {Kudoh}, {Kawamuro}, {Baba}, {Matsumoto}, {Fujita}, \&
  {Tristram}}]{2023arXiv230503993I}
{Izumi}, T., {Wada}, K., {Imanishi}, M., {et~al.} 2023, arXiv e-prints,
  arXiv:2305.03993, \dodoi{10.48550/arXiv.2305.03993}

\bibitem[{{Kawaguchi}(2003)}]{Kawaguchi2003}
{Kawaguchi}, T. 2003, \apj, 593, 69, \dodoi{10.1086/376404}

\bibitem[{{Kawaguchi} \& {Mori}(2010)}]{2010ApJ...724L.183K}
{Kawaguchi}, T., \& {Mori}, M. 2010, \apjl, 724, L183,
  \dodoi{10.1088/2041-8205/724/2/L183}

\bibitem[{{Kawaguchi} \& {Mori}(2011)}]{2011ApJ...737..105K}
---. 2011, \apj, 737, 105, \dodoi{10.1088/0004-637X/737/2/105}

\bibitem[{{Kishimoto} {et~al.}(2011{\natexlab{a}}){Kishimoto}, {H{\"o}nig},
  {Antonucci}, {Barvainis}, {Kotani}, {Tristram}, {Weigelt}, \&
  {Levin}}]{2011A&A...527A.121K}
{Kishimoto}, M., {H{\"o}nig}, S.~F., {Antonucci}, R., {et~al.}
  2011{\natexlab{a}}, \aap, 527, A121, \dodoi{10.1051/0004-6361/201016054}

\bibitem[{{Kishimoto} {et~al.}(2011{\natexlab{b}}){Kishimoto}, {H{\"o}nig},
  {Antonucci}, {Millour}, {Tristram}, \& {Weigelt}}]{2011A&A...536A..78K}
---. 2011{\natexlab{b}}, \aap, 536, A78, \dodoi{10.1051/0004-6361/201117367}

\bibitem[{{Kishimoto} {et~al.}(2007){Kishimoto}, {H{\"o}nig}, {Beckert}, \&
  {Weigelt}}]{2007A&A...476..713K}
{Kishimoto}, M., {H{\"o}nig}, S.~F., {Beckert}, T., \& {Weigelt}, G. 2007,
  \aap, 476, 713, \dodoi{10.1051/0004-6361:20077911}

\bibitem[{{Klassen} {et~al.}(2014){Klassen}, {Kuiper}, {Pudritz}, {Peters},
  {Banerjee}, \& {Buntemeyer}}]{2014ApJ...797....4K}
{Klassen}, M., {Kuiper}, R., {Pudritz}, R.~E., {et~al.} 2014, \apj, 797, 4,
  \dodoi{10.1088/0004-637X/797/1/4}

\bibitem[{{Koshida} {et~al.}(2014){Koshida}, {Minezaki}, {Yoshii}, {Kobayashi},
  {Sakata}, {Sugawara}, {Enya}, {Suganuma}, {Tomita}, {Aoki}, \&
  {Peterson}}]{2014ApJ...788..159K}
{Koshida}, S., {Minezaki}, T., {Yoshii}, Y., {et~al.} 2014, \apj, 788, 159,
  \dodoi{10.1088/0004-637X/788/2/159}

\bibitem[{{Liu} {et~al.}(2021){Liu}, {Luo}, {Brandt}, {Brotherton},
  {Gallagher}, {Ni}, {Shemmer}, \& {Timlin}}]{2021ApJ...910..103L}
{Liu}, H., {Luo}, B., {Brandt}, W.~N., {et~al.} 2021, \apj, 910, 103,
  \dodoi{10.3847/1538-4357/abe37f}

\bibitem[{{L{\'o}pez-Gonzaga} {et~al.}(2016){L{\'o}pez-Gonzaga}, {Burtscher},
  {Tristram}, {Meisenheimer}, \& {Schartmann}}]{2016A&A...591A..47L}
{L{\'o}pez-Gonzaga}, N., {Burtscher}, L., {Tristram}, K.~R.~W., {Meisenheimer},
  K., \& {Schartmann}, M. 2016, \aap, 591, A47,
  \dodoi{10.1051/0004-6361/201527590}

\bibitem[{{Lyu} {et~al.}(2022){Lyu}, {Wu}, {Yan}, {Yu}, \&
  {Liu}}]{2022ApJ...927..227L}
{Lyu}, B., {Wu}, Q., {Yan}, Z., {Yu}, W., \& {Liu}, H. 2022, \apj, 927, 227,
  \dodoi{10.3847/1538-4357/ac5256}

\bibitem[{{Lyu} \& {Rieke}(2021)}]{2021ApJ...912..126L}
{Lyu}, J., \& {Rieke}, G.~H. 2021, \apj, 912, 126,
  \dodoi{10.3847/1538-4357/abee14}

\bibitem[{{Mathis} {et~al.}(1977){Mathis}, {Rumpl}, \&
  {Nordsieck}}]{1977ApJ...217..425M}
{Mathis}, J.~S., {Rumpl}, W., \& {Nordsieck}, K.~H. 1977, \apj, 217, 425,
  \dodoi{10.1086/155591}

\bibitem[{{Matsumoto} {et~al.}(2022){Matsumoto}, {Nakagawa}, {Wada}, {Baba},
  {Onishi}, {Uzuo}, {Isobe}, \& {Kudoh}}]{2022ApJ...934...25M}
{Matsumoto}, K., {Nakagawa}, T., {Wada}, K., {et~al.} 2022, \apj, 934, 25,
  \dodoi{10.3847/1538-4357/ac755f}

\bibitem[{{Matsumoto} {et~al.}(2019){Matsumoto}, {Asahina}, {Kudoh},
  {Kawashima}, {Matsumoto}, {Takahashi}, {Minoshima}, {Zenitani}, {Miyoshi}, \&
  {Matsumoto}}]{2019PASJ...71...83M}
{Matsumoto}, Y., {Asahina}, Y., {Kudoh}, Y., {et~al.} 2019, \pasj, 71, 83,
  \dodoi{10.1093/pasj/psz064}

\bibitem[{{Meijerink} \& {Spaans}(2005)}]{2005A&A...436..397M}
{Meijerink}, R., \& {Spaans}, M. 2005, \aap, 436, 397,
  \dodoi{10.1051/0004-6361:20042398}

\bibitem[{{Minezaki} {et~al.}(2019){Minezaki}, {Yoshii}, {Kobayashi},
  {Sugawara}, {Sakata}, {Enya}, {Koshida}, {Tomita}, {Suganuma}, {Aoki}, \&
  {Peterson}}]{2019ApJ...886..150M}
{Minezaki}, T., {Yoshii}, Y., {Kobayashi}, Y., {et~al.} 2019, \apj, 886, 150,
  \dodoi{10.3847/1538-4357/ab4f7b}

\bibitem[{{Mor} \& {Netzer}(2012)}]{2012MNRAS.420..526M}
{Mor}, R., \& {Netzer}, H. 2012, \mnras, 420, 526,
  \dodoi{10.1111/j.1365-2966.2011.20060.x}

\bibitem[{{Namekata} \& {Umemura}(2016)}]{2016MNRAS.460..980N}
{Namekata}, D., \& {Umemura}, M. 2016, \mnras, 460, 980,
  \dodoi{10.1093/mnras/stw862}

\bibitem[{{Netzer}(1987)}]{1987MNRAS.225...55N}
{Netzer}, H. 1987, \mnras, 225, 55, \dodoi{10.1093/mnras/225.1.55}

\bibitem[{{Netzer}(2015)}]{2015ARA&A..53..365N}
---. 2015, \araa, 53, 365, \dodoi{10.1146/annurev-astro-082214-122302}

\bibitem[{{Noda} {et~al.}(2020){Noda}, {Kawamuro}, {Kokubo}, \&
  {Minezaki}}]{2020MNRAS.495.2921N}
{Noda}, H., {Kawamuro}, T., {Kokubo}, M., \& {Minezaki}, T. 2020, \mnras, 495,
  2921, \dodoi{10.1093/mnras/staa1376}

\bibitem[{{Ogawa} {et~al.}(2022){Ogawa}, {Ueda}, {Wada}, \&
  {Mizumoto}}]{2022ApJ...925...55O}
{Ogawa}, S., {Ueda}, Y., {Wada}, K., \& {Mizumoto}, M. 2022, \apj, 925, 55,
  \dodoi{10.3847/1538-4357/ac3cb9}

\bibitem[{{Ohsuga} {et~al.}(2005){Ohsuga}, {Mori}, {Nakamoto}, \&
  {Mineshige}}]{2005ApJ...628..368O}
{Ohsuga}, K., {Mori}, M., {Nakamoto}, T., \& {Mineshige}, S. 2005, \apj, 628,
  368, \dodoi{10.1086/430728}

\bibitem[{{Onishi} {et~al.}(2021){Onishi}, {Nakagawa}, {Baba}, {Matsumoto},
  {Isobe}, {Shirahata}, {Terada}, {Usuda}, \& {Oyabu}}]{2021ApJ...921..141O}
{Onishi}, S., {Nakagawa}, T., {Baba}, S., {et~al.} 2021, \apj, 921, 141,
  \dodoi{10.3847/1538-4357/ac1c6d}

\bibitem[{{Schartmann} {et~al.}(2011){Schartmann}, {Krause}, \&
  {Burkert}}]{2011MNRAS.415..741S}
{Schartmann}, M., {Krause}, M., \& {Burkert}, A. 2011, \mnras, 415, 741,
  \dodoi{10.1111/j.1365-2966.2011.18751.x}

\bibitem[{{Schartmann} {et~al.}(2005){Schartmann}, {Meisenheimer}, {Camenzind},
  {Wolf}, \& {Henning}}]{2005A&A...437..861S}
{Schartmann}, M., {Meisenheimer}, K., {Camenzind}, M., {Wolf}, S., \&
  {Henning}, T. 2005, \aap, 437, 861, \dodoi{10.1051/0004-6361:20042363}

\bibitem[{{Schartmann} {et~al.}(2014){Schartmann}, {Wada}, {Prieto}, {Burkert},
  \& {Tristram}}]{2014MNRAS.445.3878S}
{Schartmann}, M., {Wada}, K., {Prieto}, M.~A., {Burkert}, A., \& {Tristram},
  K.~R.~W. 2014, \mnras, 445, 3878, \dodoi{10.1093/mnras/stu2020}

\bibitem[{{Shakura} \& {Sunyaev}(1973)}]{1973A&A....24..337S}
{Shakura}, N.~I., \& {Sunyaev}, R.~A. 1973, \aap, 500, 33

\bibitem[{{Suganuma} {et~al.}(2006){Suganuma}, {Yoshii}, {Kobayashi},
  {Minezaki}, {Enya}, {Tomita}, {Aoki}, {Koshida}, \&
  {Peterson}}]{2006ApJ...639...46S}
{Suganuma}, M., {Yoshii}, Y., {Kobayashi}, Y., {et~al.} 2006, \apj, 639, 46,
  \dodoi{10.1086/499326}

\bibitem[{{Tristram} {et~al.}(2014){Tristram}, {Burtscher}, {Jaffe},
  {Meisenheimer}, {H{\"o}nig}, {Kishimoto}, {Schartmann}, \&
  {Weigelt}}]{2014A&A...563A..82T}
{Tristram}, K. R.~W., {Burtscher}, L., {Jaffe}, W., {et~al.} 2014, \aap, 563,
  A82, \dodoi{10.1051/0004-6361/201322698}

\bibitem[{{Tsai} \& {Mathews}(1995)}]{1995ApJ...448...84T}
{Tsai}, J.~C., \& {Mathews}, W.~G. 1995, \apj, 448, 84, \dodoi{10.1086/175943}

\bibitem[{{Uzuo} {et~al.}(2021){Uzuo}, {Wada}, {Izumi}, {Baba}, {Matsumoto}, \&
  {Kudoh}}]{2021ApJ...915...89U}
{Uzuo}, T., {Wada}, K., {Izumi}, T., {et~al.} 2021, \apj, 915, 89,
  \dodoi{10.3847/1538-4357/ac013d}

\bibitem[{{Wada}(2012)}]{2012ApJ...758...66W}
{Wada}, K. 2012, \apj, 758, 66, \dodoi{10.1088/0004-637X/758/1/66}

\bibitem[{{Wada}(2015)}]{2015ApJ...812...82W}
---. 2015, \apj, 812, 82, \dodoi{10.1088/0004-637X/812/1/82}

\bibitem[{{Wada} {et~al.}(2018{\natexlab{a}}){Wada}, {Fukushige}, {Izumi}, \&
  {Tomisaka}}]{2018ApJ...852...88W}
{Wada}, K., {Fukushige}, R., {Izumi}, T., \& {Tomisaka}, K. 2018{\natexlab{a}},
  \apj, 852, 88, \dodoi{10.3847/1538-4357/aa9e53}

\bibitem[{{Wada} {et~al.}(2009){Wada}, {Papadopoulos}, \&
  {Spaans}}]{2009ApJ...702...63W}
{Wada}, K., {Papadopoulos}, P.~P., \& {Spaans}, M. 2009, \apj, 702, 63,
  \dodoi{10.1088/0004-637X/702/1/63}

\bibitem[{{Wada} {et~al.}(2016){Wada}, {Schartmann}, \&
  {Meijerink}}]{2016ApJ...828L..19W}
{Wada}, K., {Schartmann}, M., \& {Meijerink}, R. 2016, \apjl, 828, L19,
  \dodoi{10.3847/2041-8205/828/2/L19}

\bibitem[{{Wada} {et~al.}(2018{\natexlab{b}}){Wada}, {Yonekura}, \&
  {Nagao}}]{2018ApJ...867...49W}
{Wada}, K., {Yonekura}, K., \& {Nagao}, T. 2018{\natexlab{b}}, \apj, 867, 49,
  \dodoi{10.3847/1538-4357/aae204}

\bibitem[{{Whalen} \& {Norman}(2006)}]{2006ApJS..162..281W}
{Whalen}, D., \& {Norman}, M.~L. 2006, \apjs, 162, 281, \dodoi{10.1086/499072}

\bibitem[{{Williamson} {et~al.}(2019){Williamson}, {H{\"o}nig}, \&
  {Venanzi}}]{2019ApJ...876..137W}
{Williamson}, D., {H{\"o}nig}, S., \& {Venanzi}, M. 2019, \apj, 876, 137,
  \dodoi{10.3847/1538-4357/ab17d5}

\bibitem[{{Williamson} {et~al.}(2020){Williamson}, {H{\"o}nig}, \&
  {Venanzi}}]{2020ApJ...897...26W}
---. 2020, \apj, 897, 26, \dodoi{10.3847/1538-4357/ab989e}

\end{thebibliography}
\bibliographystyle{aasjournal}



\end{document}